\documentclass[aps]{revtex4}
\usepackage{amssymb,epsf}

\begin{document}

\title{Geometrical method for thermal instability of nonlinearly charged BTZ
Black Holes}
\author{Seyed Hossein Hendi$^{1,2}$\footnote{
email address: hendi@shirazu.ac.ir}, Shahram Panahiyan$^{1}$\footnote{%
email address: ziexify@gmail.com} and Behzad Eslam Panah$^{1}$\footnote{%
email address: behzad$_{-}$eslampanah@yahoo.com}}
\affiliation{$^1$ Physics Department and Biruni Observatory,
College of Sciences, Shiraz University, Shiraz 71454, Iran\\
$^2$ Research Institute for Astronomy and Astrophysics of Maragha
(RIAAM), Maragha, Iran}

\begin{abstract}
In this paper we consider three dimensional BTZ black holes with
three models of nonlinear electrodynamics as source. Calculating
heat capacity, we study the stability and phase transitions of
these black holes. We show that Maxwell, logarithmic and
exponential theories yield only type one phase transition which is
related to the root(s) of heat capacity. Whereas for correction
form of nonlinear electrodynamics, heat capacity contains two
roots and one divergence point. Next, we use geometrical approach
for studying classical thermodynamical behavior of the system. We
show that Weinhold and Ruppeiner metrics fail to provide fruitful
results and the consequences of the Quevedo approach are not
completely matched to the heat capacity results. Then, we employ a
new metric for solving this problem. We show that this approach is
successful and all divergencies of its Ricci scalar and phase
transition points coincide. We also show that there is no phase
transition for uncharged BTZ black holes.
\end{abstract}

\maketitle

\section{Introduction}

One of the interesting subjects for recent gravitational studies is the
investigation of three dimensional black holes \cite%
{Astorino1,Astorino2,Astorino3,Astorino4,Astorino5,Astorino6,Astorino7,Astorino8,Astorino9}%
. Considering three dimensional solutions helps us to find a
profound insight in the black hole physics, quantum view of
gravity and also its relations to string theory
\cite{Carlip1,Carlip2,Carlip3,Witten1,Witten2}. Moreover, three
dimensional spacetimes play an essential role to improve our
understanding of gravitational interaction in low dimensional
manifolds \cite{Witten2007}. Due to these facts, some of
physicists have an interest
in the $(2+1)$-dimensional manifolds and their attractive properties \cite%
{BTZ1,BTZ2,Nojiri19981,Nojiri19982,Nojiri19983,Nojiri19984,Nojiri19985,Nojiri19986,Nojiri19987,Nojiri19988,Nojiri19989,Nojiri199810}%
.

The Maxwell theory is in agreement with experimental results, but
it fails regarding some important issues such as self energy of
point-like charges which motivates one to regard nonlinear
electrodynamics (NED). There are some evidences that motivate one
to consider NED theories: solving the problem of point-like charge
self energy, understanding the nature of different complex
systems, obtaining more information and insight regarding to
quantum gravity, compatible with AdS/CFT correspondence and string
theory frames, description of pair creation for Hawking radiation
and the behavior of the
compact astrophysical objects such as neutron stars and pulsars \cite%
{Chen,Fukuma,Aros}. Therefore, many authors investigated the black hole
solutions with nonlinear sources \cite%
{nonsources1,nonsources2,nonsources3,nonsources4,nonsources5,nonsources6,nonsources7,nonsources8,nonsources9,nonsources10,nonsources11,nonsources13,nonsources14,nonsources15,nonsources16,nonsources17,nonsources18,nonsources19,nonsources20}%
.

On the other hand, thermodynamical structure of the black holes,
has been of great interest. It is due to the fact that, according
to AdS/CFT correspondence, black hole thermodynamics provides a
machinery to map a solution in AdS spacetime to a conformal field
on the boundary of this spacetime
\cite{Witten1,Witten2,Witten2007}. Also, it was recently pointed
out that considering cosmological constant as a thermodynamical
variable leads to the behavior similar to the Van Der Waals liquid/gas system \cite%
{PVpaper1,PVpaper2,PVpaper3,PVpaper4,PVpaper5,PVpaper6,PVpaper7,PVpaper8}.
In addition, phase transition of the black holes plays an
important role in exploring the critical behavior of the system
near critical points. There are several approaches that one can
employ to study the phase transition. One of these approaches is
studying the behavior of the heat capacity. It is argued that
roots and divergence points of the heat capacity are representing
two types of phase transition \cite{PT1,PT2,PT3,PT4,PT5,PT6,PT7}.
In addition, studying the heat capacity and its behavior, enable
one to study the thermal stability of the black holes \cite
{Stability1,Stability2,Stability3,Stability4,Correction2,Mamasani}
.

Another approach for studying phase transition of black holes is
through thermodynamical geometry. The concept is to construct a
spacetime by employing the thermodynamical properties of the
system. Then, by studying the divergence points of thermodynamical
Ricci scalar of the metric, one can investigate phase transition
points. In other words, it is expected that divergencies of
thermodynamical Ricci scalar (TRS) coincide with phase transition
points of the black holes. Firstly, Weinhold introduced
differential geometric concepts into ordinary thermodynamics
\cite{Weinhold1,Weinhold2}. He considered a kind of metric defined
as the second derivatives of internal energy with respect to
entropy and other extensive quantities for a thermodynamical
system. Later Ruppeiner \cite{Ruppeiner1,Ruppeiner2} introduced
another metric and defined the minus second derivatives of entropy
with respect to the internal energy and other extensive
quantities. The Ruppeiner metric is conformal to the Weinhold
metric with the inverse temperature as the conformal factor. It is
notable that, both metrics have been applied to study the
thermodynamical geometry
of ordinary systems \cite%
{Janyszek1,Janyszek2,Janyszek3,Janyszek4,Janyszek5,Janyszek6}. In
particular, it was found that the Ruppeiner geometry carries information of
phase structure of thermodynamical system. For the systems with no
statistical mechanical interactions (for example, ideal gas), the scalar
curvature is zero and the Ruppeiner metric is flat. Because of the success
of their applications to ordinary thermodynamical systems, they have also
been used to study black hole phase structures and lots of results have been
obtained for different types of black holes \cite%
{Ferrara1,Ferrara2,Ferrara3,Ferrara4,Ferrara5,Ferrara6}.

It is notable that these two approaches fail in order to describe phase
transition of several black holes \cite{HPEM}. In order to overcome this
problem, Quevedo proposed new types of metrics for studying geometrical
structure of the black hole thermodynamics \cite{Quevedo1,Quevedo2}. This
method was employed to study the geometrical structure of the phase
transition of the black holes \cite%
{PT1,PT2,PT3,PT4,PT5,PT6,PT7,QuevedoP1,QuevedoP2,QuevedoP3,QuevedoP4,QuevedoP5}
and proved to be a strong machinery for describing phase
transition of the black holes. But this approach was not
completely coincided with the results of classical thermodynamics
arisen from the heat capacity \cite{HPEM}. In Ref. \cite{HPEM}, a
new metric was proposed in which the denominator of its Ricci
scalar is only constructed of numerator and denominator of the
heat capacity. Several phase transition of the black holes have
been studied in context of the HPEM (Hendi-Panahiyan-Eslam
Panah-Momennia) metric \cite{HPEM}.

In this paper we study thermal stability and phase transition of
the BTZ black holes in presence of several NED models in context
of heat capacity. Then, we employ Weinhold, Ruppeiner and Quevedo
methods for studying geometrothermodynamics of these black holes.
We will see that Weinhold and Ruppeiner metrics fail to provide
fruitful results and the consequences of the Quevedo approach are
not completely matched to the heat capacity results. Then, we
employ the HPEM metric and study the phase transition of these
black holes in context of geometrothermodynamics. We end the paper
with some closing remarks.

\section{Nonlinearly charged BTZ Black hole solutions\label{Sol}}

The $(2+1)$-dimensional action of Einstein gravity with NED field in the
presence of cosmological constant is given by
\begin{equation}
I=-\frac{1}{16\pi }\int d^{3}x\sqrt{-g}\left[ R-2\Lambda +L(\mathcal{F})%
\right] ,  \label{Action}
\end{equation}%
where $R$ is the Ricci scalar, the cosmological constant is
$\Lambda =-l^{-2} $ in which $l$ is a scale factor. Also, $L(F)$
is the Lagrangian of NED, in which we consider three models. First
model was proposed by Hendi (HNED) \cite{Hendi}, second one is
Soleng theory (SNED) \cite{Soleng} and third one is correction
form of NED (CNED) \cite{Correction2}
\begin{equation}
L(\mathcal{F})=\left\{
\begin{array}{cc}
\beta ^{2}\left( \exp \left( -\frac{\mathcal{F}}{\beta ^{2}}\right)
-1\right) , & HNED\vspace{0.3cm} \\
-8\beta ^{2}\ln \left( 1+\frac{\mathcal{F}}{8\beta ^{2}}\right) , & SNED%
\vspace{0.3cm} \\
-\mathcal{F}+\alpha \mathcal{F}^{2}+\mathcal{O}(\alpha ^{2}), & CNED%
\end{array}%
\right. ,
\end{equation}%
where $\beta $ and $\alpha$ are called the nonlinearity parameters, the
Maxwell invariant $\mathcal{F}=F_{\mu \nu }F^{\mu \nu }$ in which $F_{\mu
\nu }=\partial _{\mu }A_{\nu }-\partial _{\nu }A_{\mu }$ is the
electromagnetic field tensor and $A_{\mu }$\ is the gauge potential. We
should note that for $\beta \longrightarrow \infty$ (HNED and SNED branches)
and $\alpha \longrightarrow 0$ (CNED branch) the Maxwell Lagrangian can be
recovered.

The nonlinearly charged static black hole solutions can be introduced with
the following line element
\begin{equation}
ds^{2}=-f(r)dt^{2}+\frac{dr^{2}}{f(r)}+r^{2}d\theta ^{2},  \label{Metric}
\end{equation}%
where the metric function $f(r)$ was obtained in Refs. \cite{Mamasani,Hendi}
\begin{equation}
f(r)=\frac{r^{2}}{l^{2}}-m+\left\{
\begin{array}{cc}
\frac{\beta rq\left( 1-2L_{W}\right) }{\sqrt{L_{W}}}-\frac{\beta ^{2}r^{2}}{2%
}+q^{2}\left[ \ln \left( \frac{\beta ^{2}l^{2}}{2q^{2}}\right) -{Ei}\left( 1,%
\frac{L_{W}}{2}\right) -\gamma +3\right] , & HNED\vspace{0.3cm} \\
4\beta ^{2}r^{2}\left[ \ln \left( \frac{\Gamma +1}{2}\right) +3\right] -q^{2}%
\left[ \ln \left( \frac{\beta ^{2}r^{4}\left( \Gamma -1\right) \left( \Gamma
+1\right) ^{3}}{4q^{2}l^{2}}\right) +\frac{6}{\Gamma -1}-2\right] , & SNED%
\vspace{0.3cm} \\
-2q^{2}\ln \left( \frac{r}{l}\right) -\frac{2\alpha q^{4}}{r^{2}}+\mathcal{O}%
(\alpha ^{2}), & CNED%
\end{array}%
\right. ,  \label{metric3dim}
\end{equation}%
where $\Gamma =\sqrt{1+\frac{q^{2}}{r^{2}\beta ^{2}}}$, $m$ and $q$ are
integrations constant which are related to mass parameter and the electric
charge of the black hole, respectively. In addition, $L_{W}=LambertW\left(
\frac{4q^{2}}{\beta ^{2}r^{2}}\right) $, $\gamma =\gamma (0)\simeq 0.57722$
and the special function ${Ei}\left( 1,x\right) =\int_{1}^{\infty }\frac{%
e^{-xz}}{z}dz$.

The entropy and the electric charge of the obtained NED black hole solutions
can be calculated with the following forms \cite{Mamasani,Hendi}
\begin{eqnarray}
S &=&\frac{\pi r_{+}}{2},  \label{Entropy} \\
Q &=&\frac{\pi q}{2},  \label{Q}
\end{eqnarray}%
where $r_{+}$ is the event horizon of black hole. On the other hand, the
quasi-local mass, which is related to geometrical mass, can be obtained as
\cite{Mamasani,Hendi}
\begin{equation}
M=\frac{\pi }{8}m.  \label{Mass}
\end{equation}

Regarding Eqs. (\ref{Entropy}) and (\ref{Q}) with (\ref{Mass}) and
obtaining $m$ by using $f(r=r_{+})=0$, for these three cases of
BTZ black holes we can write
\begin{equation}
M=\frac{1}{8\pi}\times \left\{
\begin{array}{cc}
\frac{4\beta SQ\left( 1-2L_{W}^{\prime }\right) }{\sqrt{L_{W}^{\prime }}}-%
\frac{2S^{2}\left( l^{2}\beta ^{2}-2\right) }{l^{2}}+4Q^{2}\left[ \ln \left(
\frac{\beta ^{2}l^{2}\pi ^{2}}{Q^{2}}\right) -{Ei}\left( 1,\frac{%
L_{W}^{\prime }}{2}\right) -\gamma +3\right] , & HNED\vspace{0.4cm} \\
\frac{4S^{2}}{l^{2}}+16\beta ^{2}S^{2}\left[ \ln \left( \frac{\Gamma
^{\prime }+1}{2}\right) +3\right] -4Q^{2}\left[ \ln \left( \frac{\beta
^{2}S^{4}\left( \Gamma ^{\prime }-1\right) \left( \Gamma ^{\prime }+1\right)
^{3}}{Q^{2}l^{2}\pi ^{2}}\right) +\frac{6}{\Gamma ^{\prime }-1}-2\right] , &
SNED\vspace{0.4cm} \\
\frac{4S^{2}}{l^{2}}-8Q^{2}\ln \left( \frac{2S}{l\pi }\right) -\frac{8Q^{4}}{%
S^{2}}\alpha , & CNED%
\end{array}%
\right. ,  \label{Mass2}
\end{equation}
where $\Gamma ^{\prime }=\sqrt{1+\frac{Q^{2}}{S^{2}\beta ^{2}}}$
and $L_{W}^{\prime }=LambertW\left( \frac{4Q^{2}}{\beta ^{2}S^{2}}
\right)$. Having conserved and thermodynamic quantities at hand,
it was shown that the first law of thermodynamics may be satisfied
\cite{Hendi}. The main goal of this paper is investigating thermal
stability and phase transition for these black holes.

\subsection{Heat capacity}

In order to investigate thermal stability and phase transition,
one can usually adopt two different approaches to the matter at
hand. In one method, the electric charge is considered as a fixed
parameter and heat capacity of the black hole will be calculated.
The positivity of the heat capacity is sufficient to ensure the
local thermal stability of the solutions and its divergencies are
corresponding to the phase transition points. This approach is
known as canonical ensemble. Another approach for studying thermal
stability of the black holes is grand canonical ensemble. In this
approach, thermal stability is investigated by calculating the
determinant of Hessian matrix of M(S,Q) with respect to its
extensive variables. The positivity of this determinant also
represents the local stability of the solutions. Although these
two approaches are different fundamentally, one expects that the
results being the same for both ensembles; i.e. ensemble
independent. Here we use the first method for studying thermal
stability. For this purpose, the system is considered to be in
fixed charge and the heat capacity has the following form
\begin{equation}
C_{Q}=\left( \frac{\partial M}{\partial S}\right) _{Q}\left( \frac{\partial
^{2}M}{\partial S^{2}}\right) _{Q}^{-1}.  \label{heat}
\end{equation}

It is notable that, when we study heat capacity for investigating the phase
transition, we encounter with two different phenomena. In one, the changes
in the signature of the heat capacity is representing a phase transition of
the system. In other words, if the heat capacity is negative, then the
system is in thermally unstable phase, whereas for the case of the positive $%
C_{Q}$, the system is thermally stable. The roots of the heat capacity in
this case are representing phase transition points which means one should
solve the following equation
\begin{equation}
\left( \frac{\partial M}{\partial S}\right) _{Q}=0,  \label{num1}
\end{equation}%
where hereafter we call this type of the phase transition as type
one. It is a matter of calculation to show that
\begin{equation}
\left( \frac{\partial M}{\partial S}\right) _{Q}=\left\{
\begin{array}{cc}
-\frac{2l^{2}Q^{2}e^{\frac{1}{2}L_{W}^{\prime }}\sqrt{L_{W}^{\prime }}+S%
\left[ S\left( \beta ^{2}l^{2}-2\right) \sqrt{L_{W}^{\prime }}\left(
1+L_{W}^{\prime 2}\right) +Q\beta l^{2}L_{W}^{\prime }\left( 2L_{W}^{\prime
}-1\right) +2Q\beta l^{2}\right] }{2\pi Sl^{2}\sqrt{L_{W}^{\prime }}\left(
1+L_{W}^{\prime }\right) }, & HNED\vspace{0.4cm} \\
\begin{array}{c}
\frac{4\beta ^{2}l^{2}Q^{2}\left( Q^{2}-S^{2}\beta ^{2}\right) \ln \left(
\frac{1+\Gamma ^{\prime }}{2}\right) +8S^{2}Q^{2}l^{2}\beta ^{4}+Q^{4}\left(
1+S^{2}+8\beta ^{2}l^{2}\right) }{2\pi S^{3}\beta ^{4}l^{2}\Gamma ^{\prime
}\left( 1-\Gamma ^{\prime }\right) \left( 1-\Gamma ^{\prime 2}\right) }+ \vspace{0.25cm}\\
\frac{S^{4}\beta ^{2}\left( 1-S^{2}\sqrt{\Gamma ^{\prime }}\right)
-6Q^{4}l^{2}}{2\pi S^{3}\beta ^{2}l^{2}\left( 1-\Gamma ^{\prime
}\right) \left( 1-\Gamma ^{\prime 2}\right) }+\frac{2S^{2}\ln
\left( \frac{\Gamma ^{\prime }+1}{2}\right) +6\beta
^{2}+Q^{2}}{\pi S\left( 1-\Gamma ^{\prime
}\right) },%
\end{array}
& SNED\vspace{0.4cm} \\
\frac{\left( 2Q^{4}\alpha -S^{2}Q^{2}\right) l^{2}+S^{4}}{\pi l^{2}S^{2}}, &
CNED%
\end{array}%
\right. , \label{dMdS}
\end{equation}%

The other case of phase transition is the divergency of the heat capacity.
In other words, the singular points of the heat capacity are representing
places in which system goes under phase transition. This assumption leads to
the fact that the roots of the denominator of the heat capacity are
representing phase transitions. Therefore, we have the following relation
for this type of phase transition
\begin{equation}
\left( \frac{\partial ^{2}M}{\partial S^{2}}\right) _{Q}=0,  \label{den1}
\end{equation}%
where we call this type of the phase transition as type two. Due
to economical reasons, we did not write the explicit relations of
$\left( \frac{\partial ^{2}M}{\partial S^{2}}\right) _{Q}$ for
different BTZ solutions.

\subsection{Geometrical Thermodynamics}

In order to have an effective geometrical approach for studying
phase transition of a system, one can build a suitable
thermodynamical metric and investigate its Ricci scalar.
Thermodynamical metrics were introduced based on the Hessian
matrix of the mass (internal energy) with respect to the extensive
variables. Therefore, although the electric charge is a fixed
parameter for calculating the heat capacity in canonical ensemble,
it may be an extensive variable for constructing thermodynamical
metrics. In this method we expect that TRS diverges in both types
of the mentioned phase transition points. In other words, the
denominator of TRS must be constructed in a way that contains
roots of the denominator and numerator of the heat capacity. In
what follows, we will study the denominator of TRS of the several
geometrical approaches and follow the recently proposed
thermodynamical metric which its denominator only contains
numerator and denominator of the heat capacity, and therefore,
divergencies of TRS coincide with roots and divergences of the
heat capacity.

In order to find the roots and divergence points of heat capacity, we should
solve its numerator and denominator, separately. Solving the mentioned
equations with respect to entropy, leads to
\begin{equation}
S_{0}\equiv \left. S\right\vert _{C_{Q}=0}=\left\{
\begin{array}{cc}
\frac{4Q\beta l^{2}L\varpi }{\left( 2-\beta ^{2}l^{2}\right) \sqrt{%
1+2L\varpi }}, & HNED\vspace{0.3cm} \\
\frac{Q}{\beta \sqrt{L\omega \left( 2+L\omega \right) }}, & SNED\vspace{0.3cm%
} \\
\frac{Q}{2}\sqrt{\pm 2l\left( \sqrt{l^{2}-8\alpha }\pm l\right) }, & CNED%
\end{array}%
\right.  \label{CQ0}
\end{equation}%
and%
\begin{equation}
S_{\infty }\equiv \left. S\right\vert _{C_{Q}\longrightarrow \infty
}=\left\{
\begin{array}{cc}
\frac{2Q\beta l^{2}}{\left( \beta ^{2}l^{2}-2\right) \sqrt{2\ln \left( 1-%
\frac{2}{\beta ^{2}l^{2}}\right) }}, & HNED\vspace{0.3cm} \\
\frac{Q}{2\beta \sqrt{\exp \left( -\frac{1}{4\beta ^{2}l^{2}}\right) \left[
\exp \left( -\frac{1}{4\beta ^{2}l^{2}}\right) -1\right] }}, & SNED\vspace{%
0.3cm} \\
\frac{Q}{2}\sqrt{2l\left( \sqrt{l^{2}+24\alpha }-l\right) }, & CNED%
\end{array}%
\right.  \label{CQInf}
\end{equation}%
where%
\begin{equation}
L\varpi =LambertW\left[ -\frac{\left( \beta ^{2}l^{2}-2\right) }{2\beta
^{2}l^{2}e^{\frac{1}{2}}}\right] ,
\end{equation}%
\begin{equation}
L\omega =LambertW\left[ -2\exp \left( -\frac{8\beta ^{2}l^{2}+1}{4\beta
^{2}l^{2}}\right) \right] .
\end{equation}

It is evident from obtained equation for HNED and SNED that there is only
one real positive entropy in which heat capacity vanishes. Interestingly, in
case of CNED theory we find two roots for heat capacity. It is evident that
the roots are increasing functions of the electric charge in these theories.
As for nonlinearity parameter, in case of the HNED and SNED theories, the
root is an increasing function of $\beta $. Whereas for CNED theory, the
smaller root is an increasing function of $\alpha $ whereas the larger root
is a decreasing function of it.

Now we are in position to study the existence of the type two phase
transition point which is related to divergency of the heat capacity.
Considering HNED branch of Eq. (\ref{CQInf}), one finds $S_{\infty }$ is not
real for all values of $l$ and $\beta $. Therefore, there is no physical
divergence point for the heat capacity of HNED model. Next, for the case of
SNED model the same behavior is seen. In other words, since $0\leq \exp (%
\frac{-1}{x})<1$ for $0\leq x<\infty $, we can not obtain real $S_{\infty }$
for all values of $l$ and $\beta $. Next, we should investigate CNED model.
Regarding Eq. (\ref{CQInf}), we find that there is a divergence point for
heat capacity (we should note that in this paper we consider positive $%
\alpha $). In other words, this theory of nonlinear electromagnetic field
enjoys the phase transition of type two. The divergence point is an
increasing function of the nonlinearity and electric charge parameters. It
is worthwhile to mention that for the case of vanishing nonlinearity
parameter in this theory, $S_{\infty }$ goes to zero. In other words, there
is no divergence point for heat capacity of Maxwell theory. It is also clear
that in case of chargeless BTZ black holes, there is no root and divergence
point for heat capacity. In other words, in case of chargeless BTZ black
holes, there is not any kind of phase transition.

\begin{figure}[tbp]
$%
\begin{array}{cc}
\epsfxsize=7cm \epsffile{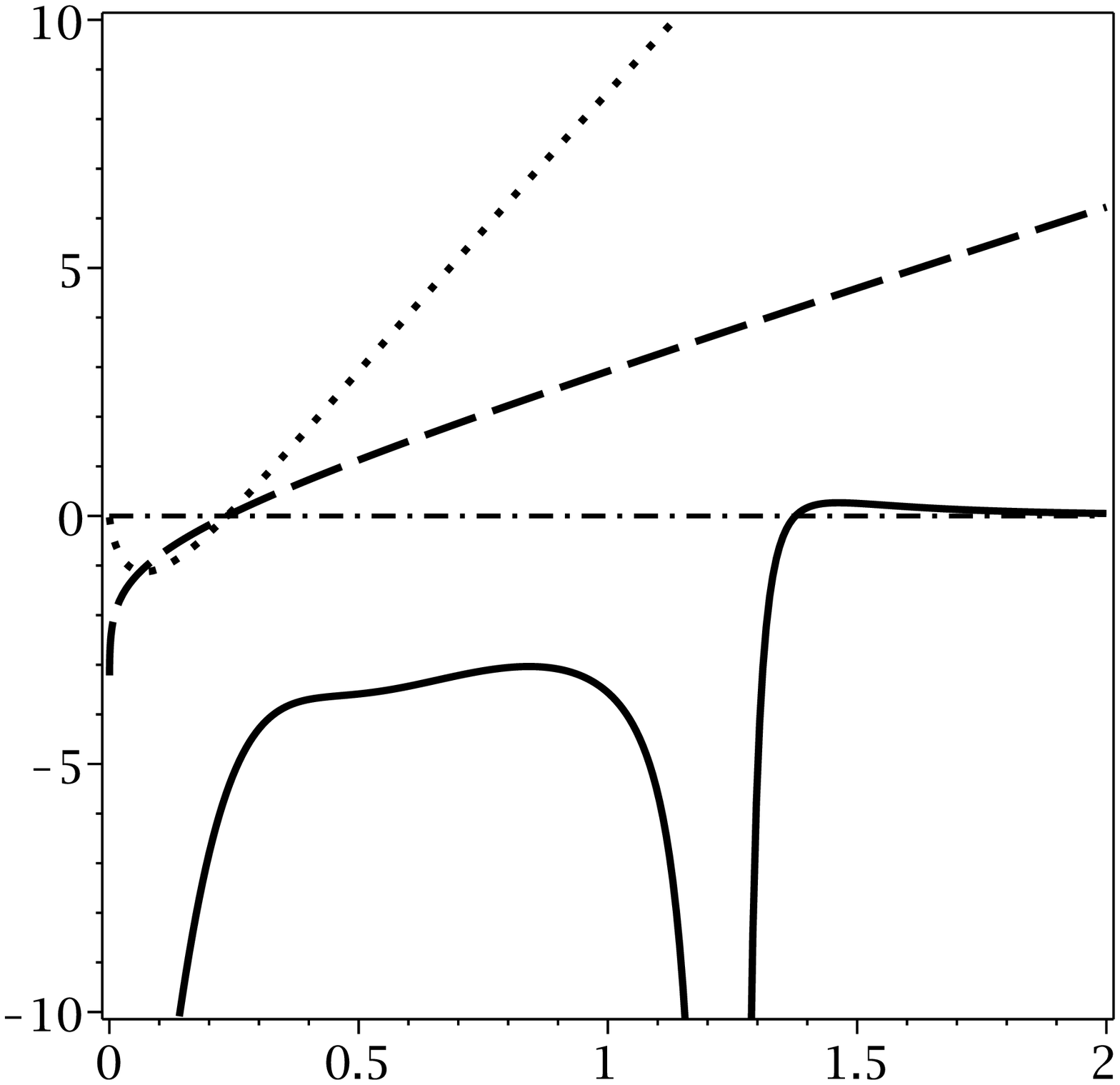} & \epsfxsize=7cm \epsffile{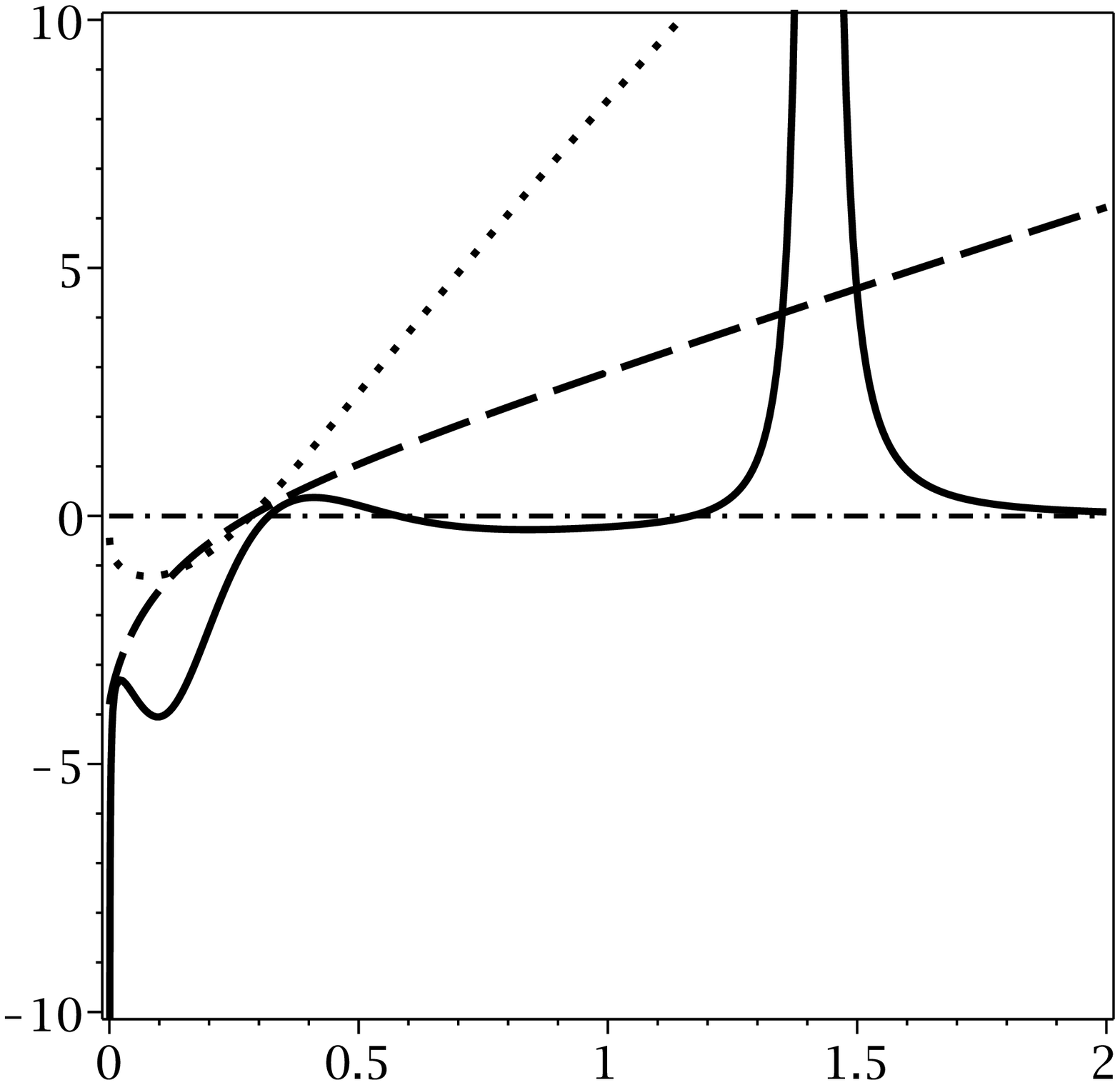} \\
\epsfxsize=7cm \epsffile{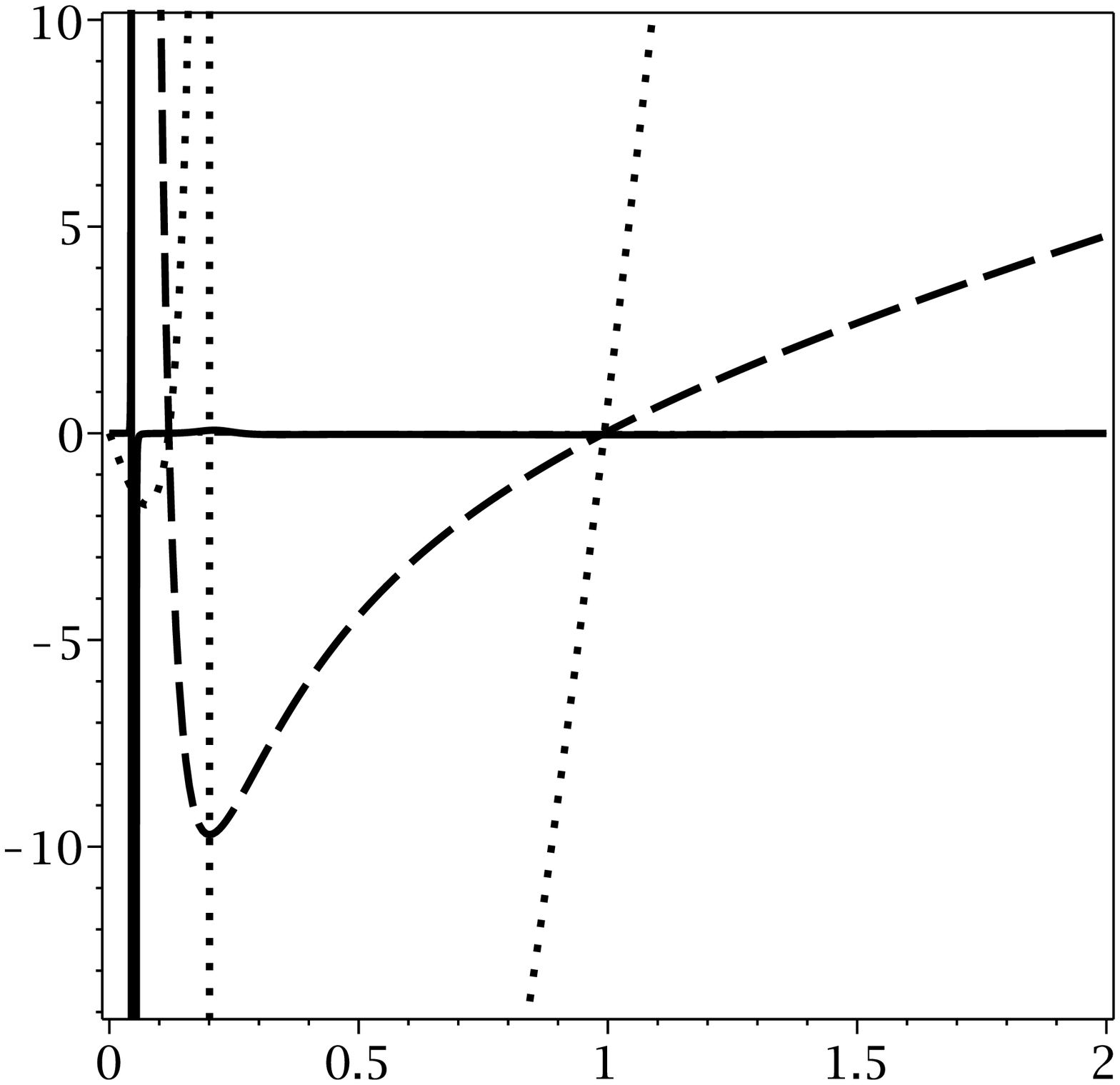} & \epsfxsize=7cm \epsffile{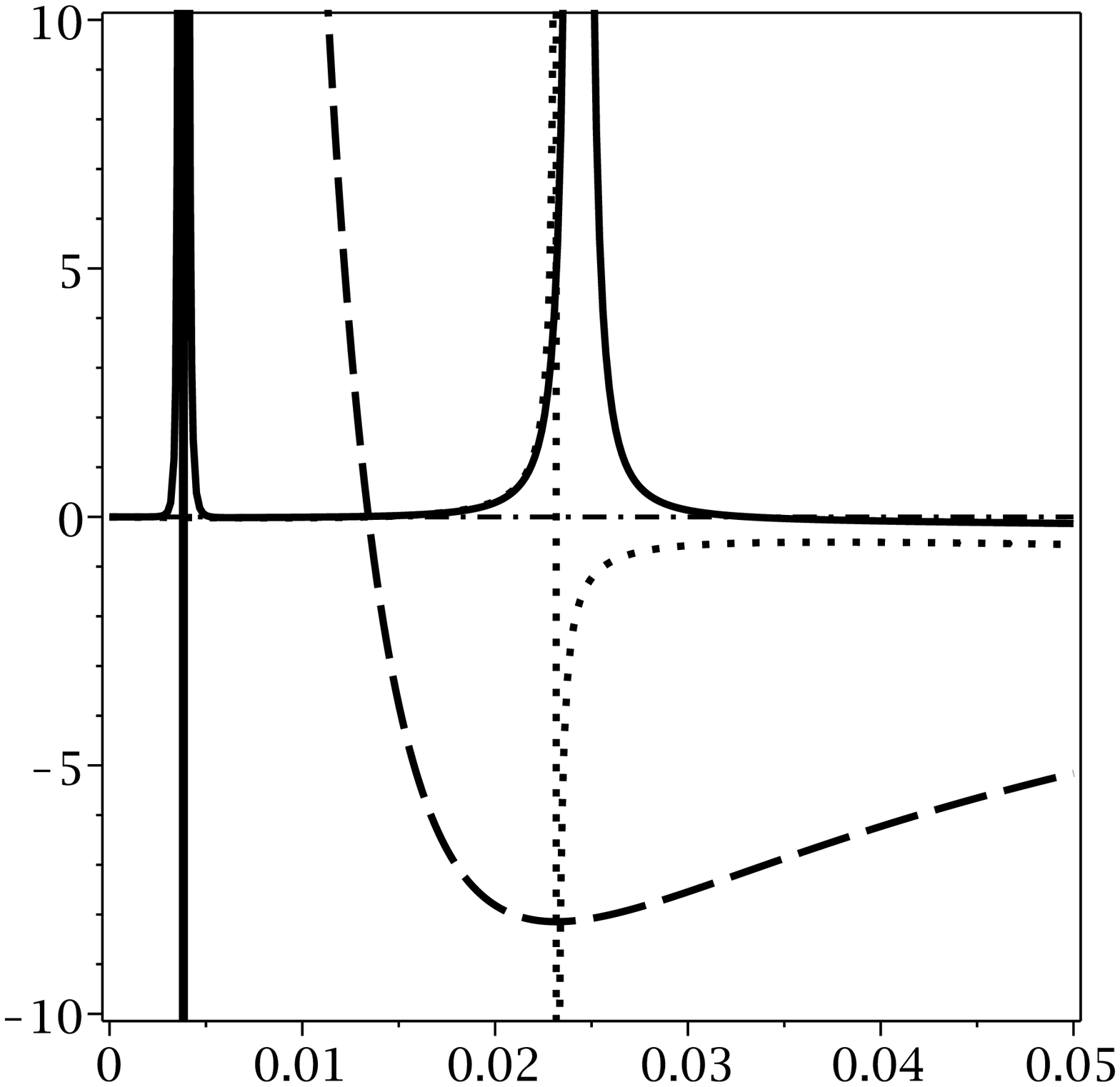}%
\end{array}
$%
\caption{Weinhold Ricci scalar (solid line), heat capacity (dotted line) and
temperature (dashed line) versus $S$ for $l=1$. \newline
\textbf{HNED model:} (up-left panel) and \textbf{SNED model:} (up-right
panel): $q=0.3$, $\protect\beta =1$. \newline
\textbf{CNED model:} (down-left panel) and \textbf{CNED model:} (down-right
panel): $q=1$, $\protect\alpha=0.007$ (different scales). }
\label{Fig1}
\end{figure}

\begin{figure}[tbp]
$%
\begin{array}{cc}
\epsfxsize=7cm \epsffile{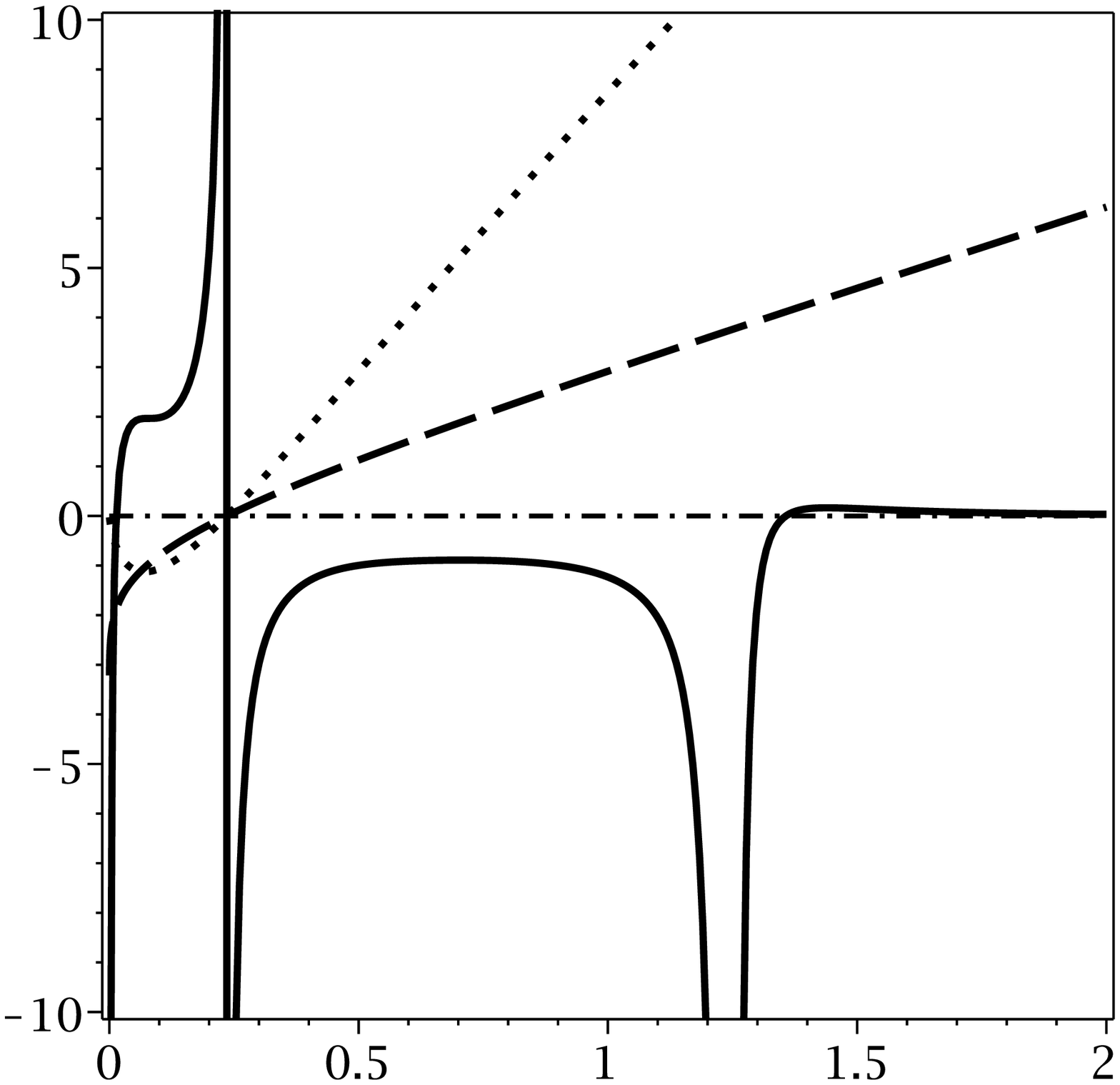} & \epsfxsize=7cm \epsffile{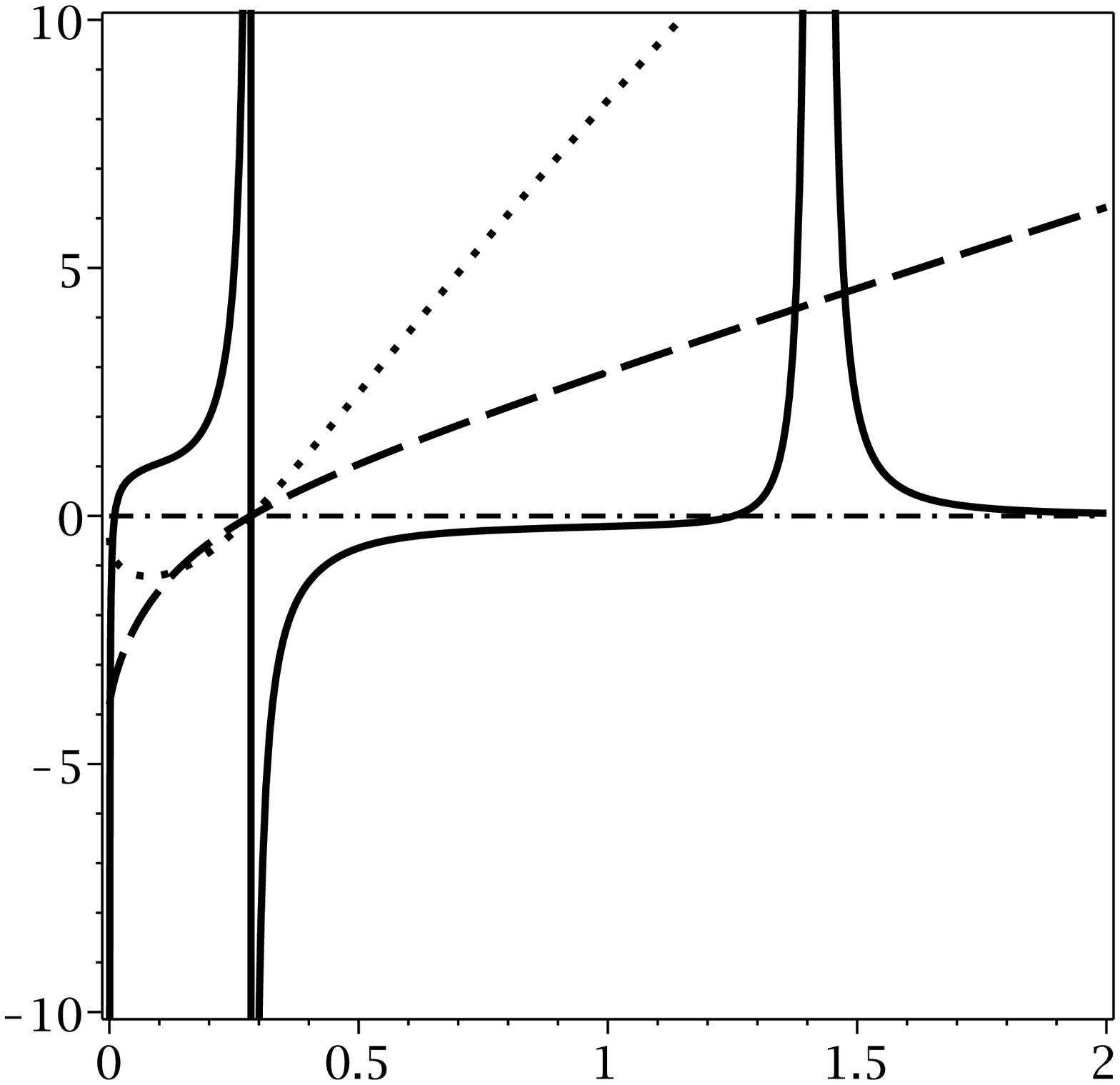} \\
\epsfxsize=7cm \epsffile{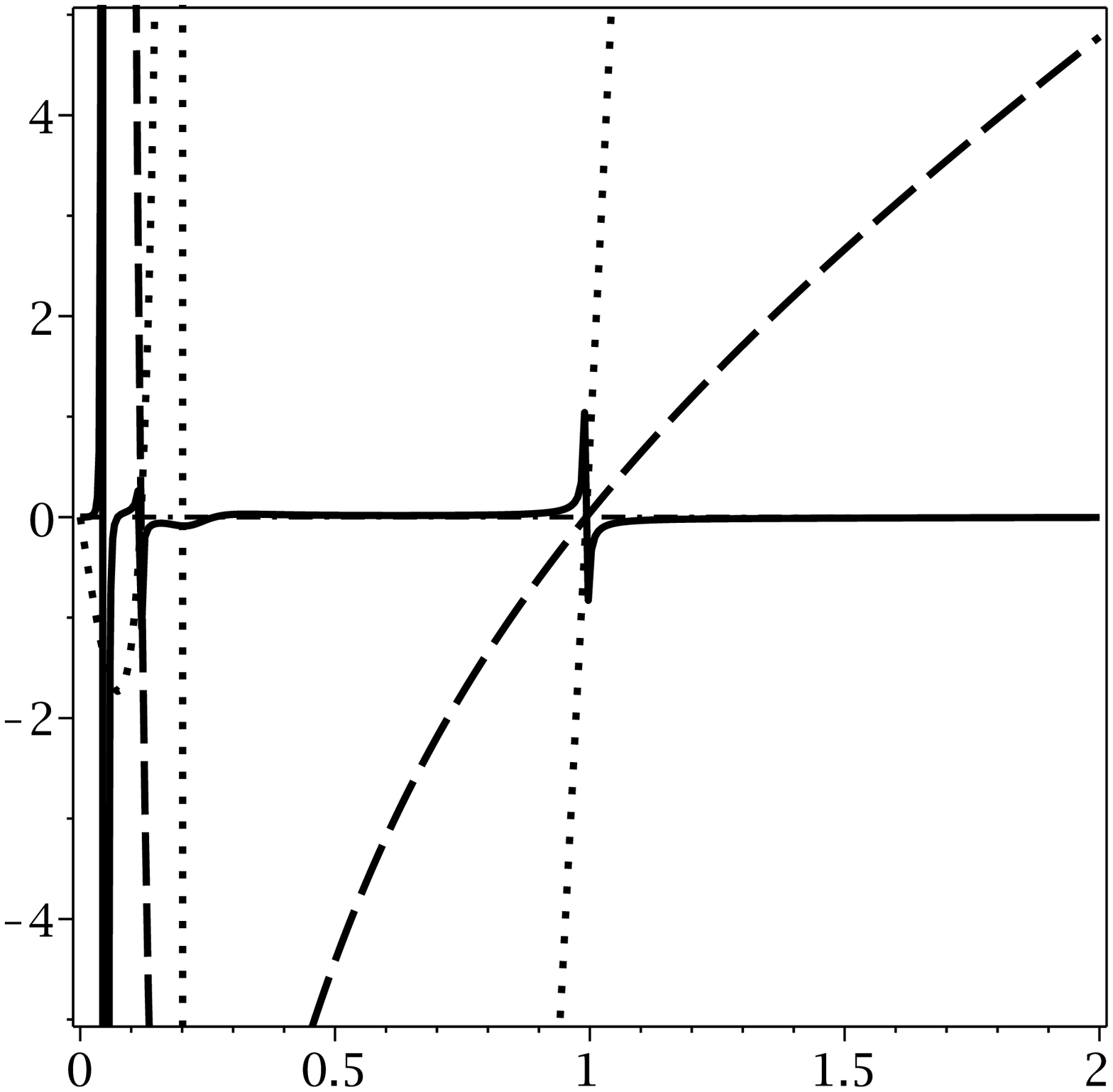} & \epsfxsize=7cm \epsffile{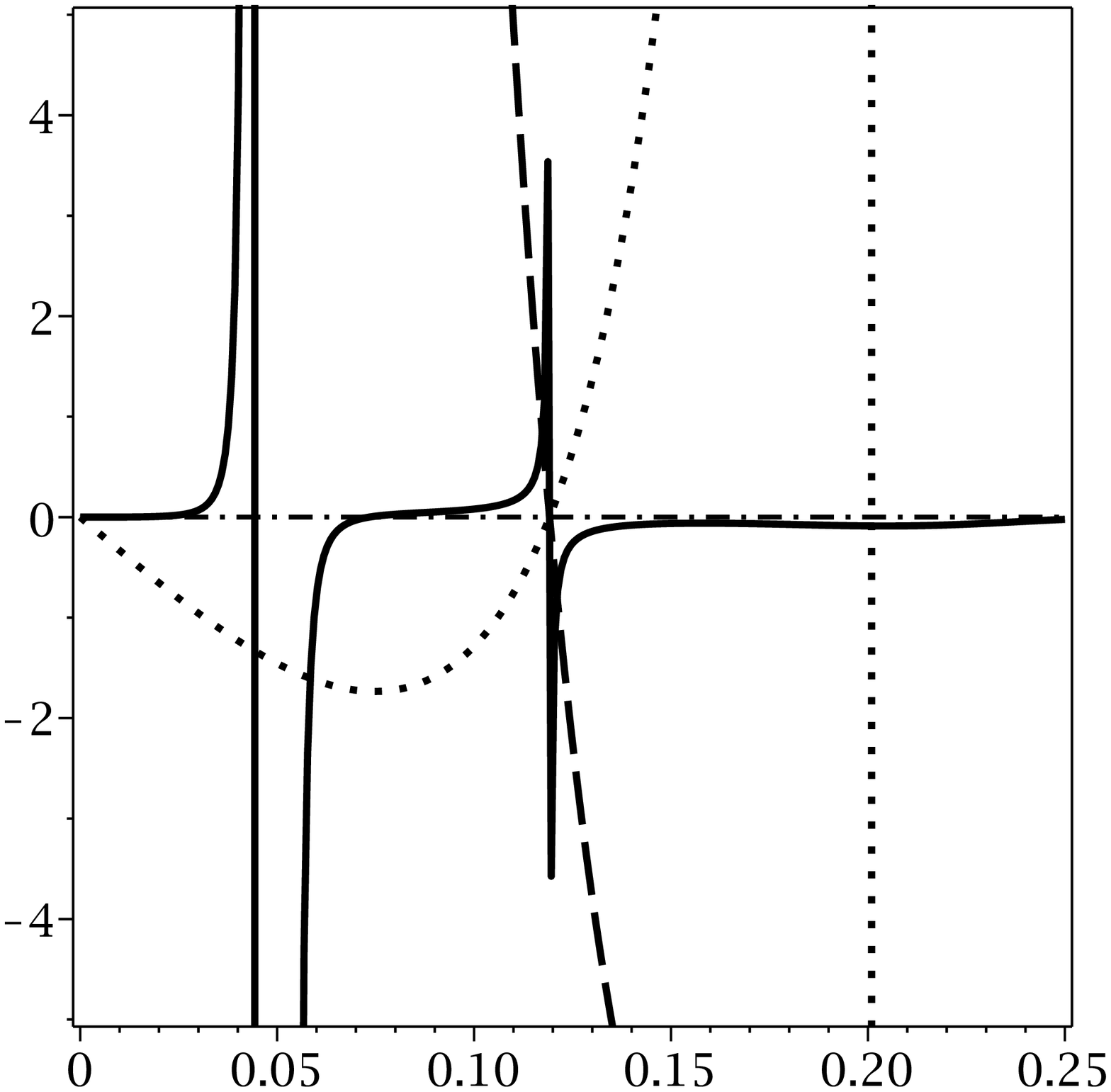}%
\end{array}
$%
\caption{Ruppeiner Ricci scalar (solid line), heat capacity (dotted line)
and temperature (dashed line) versus $S$ for $l=1$. \newline
\textbf{HNED model:} (up-left panel) and \textbf{SNED model:} (up-right
panel): $q=0.3$, $\protect\beta =1$. \newline
\textbf{CNED model:} (down-left panel) and \textbf{CNED model:} (down-right
panel): $q=1$, $\protect\alpha=0.007$ (different scales). }
\label{Fig2}
\end{figure}

\begin{figure}[tbp]
$%
\begin{array}{cc}
\epsfxsize=7cm \epsffile{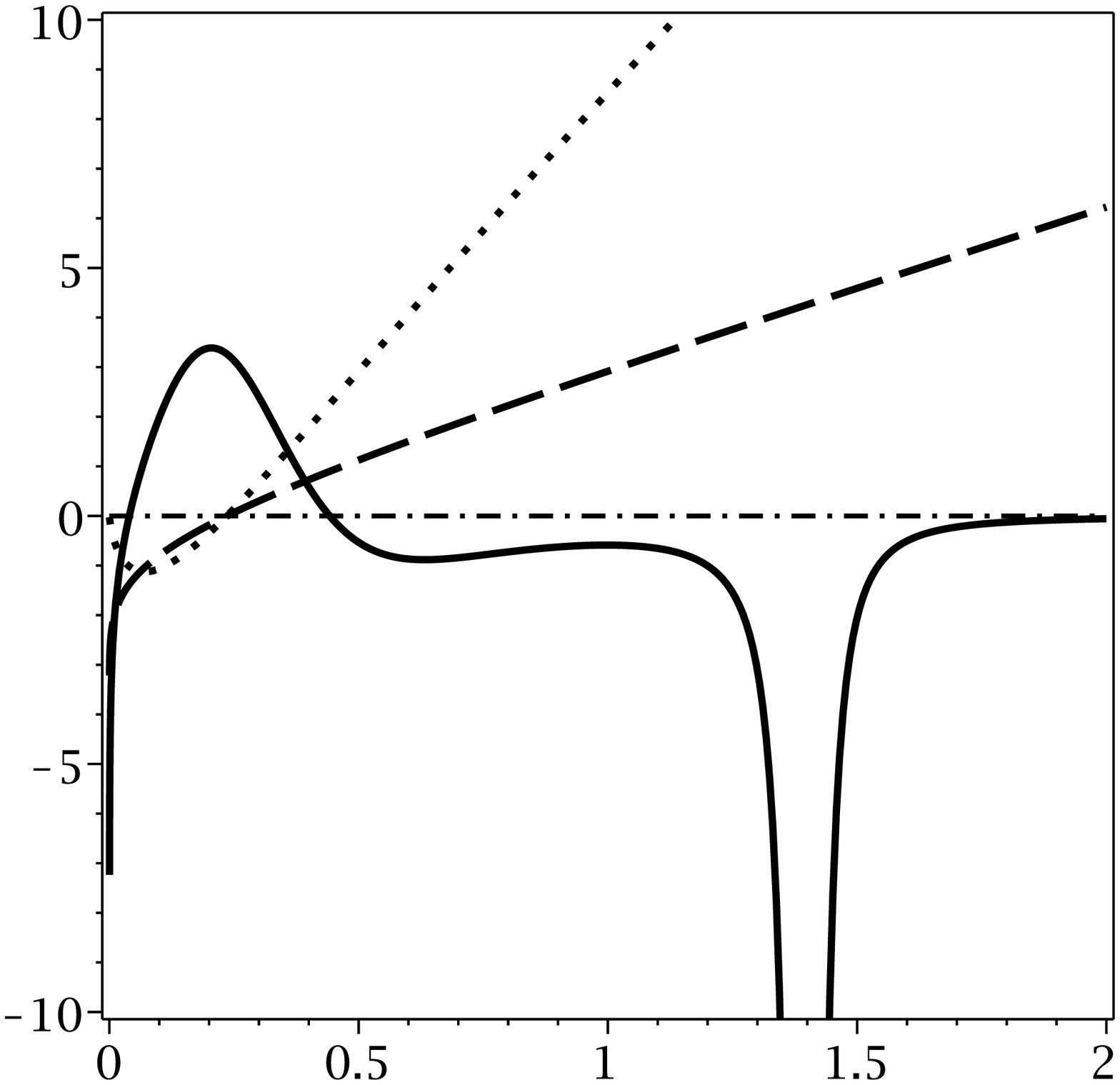} & \epsfxsize=7cm \epsffile{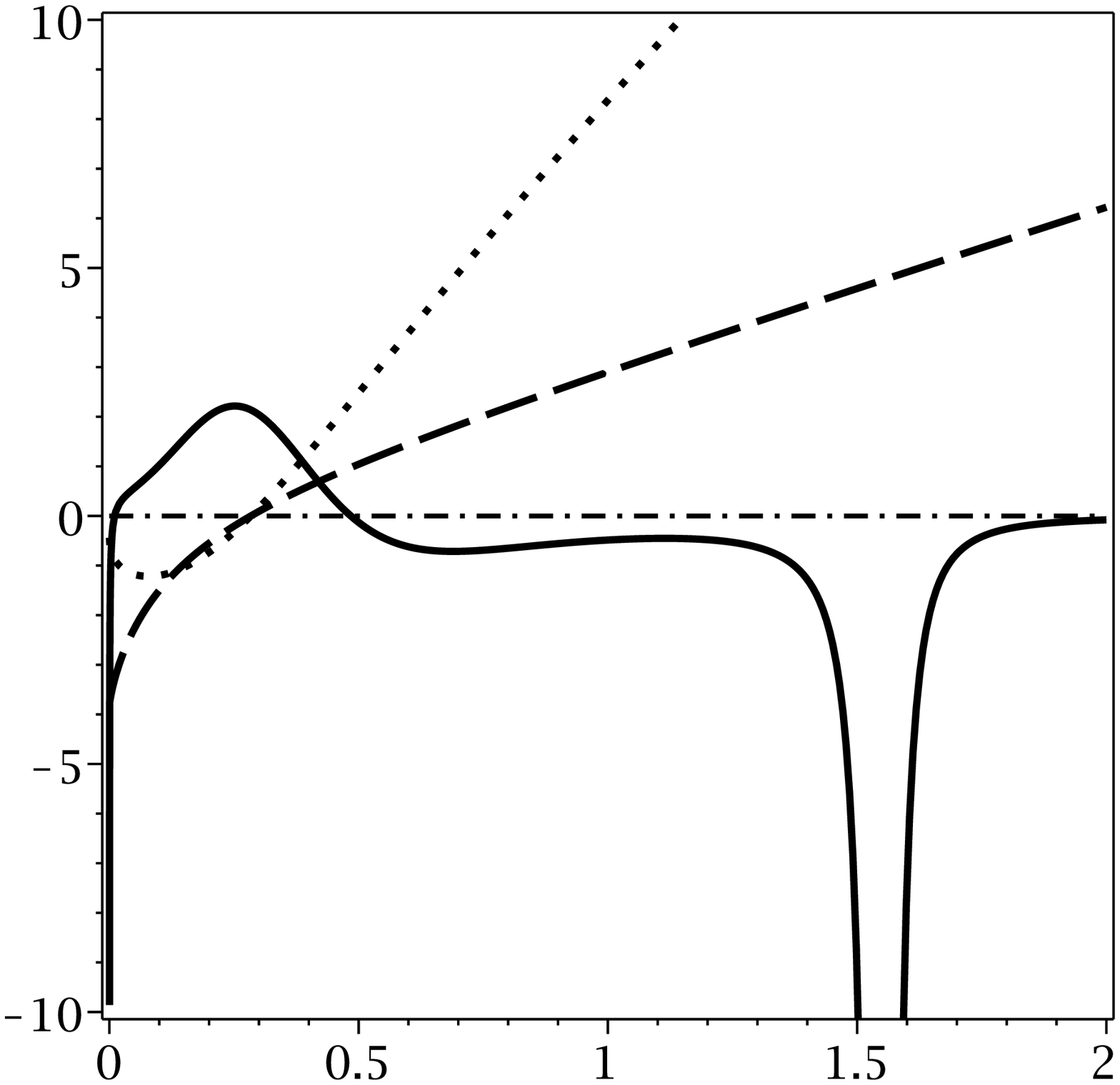} \\
\epsfxsize=7cm \epsffile{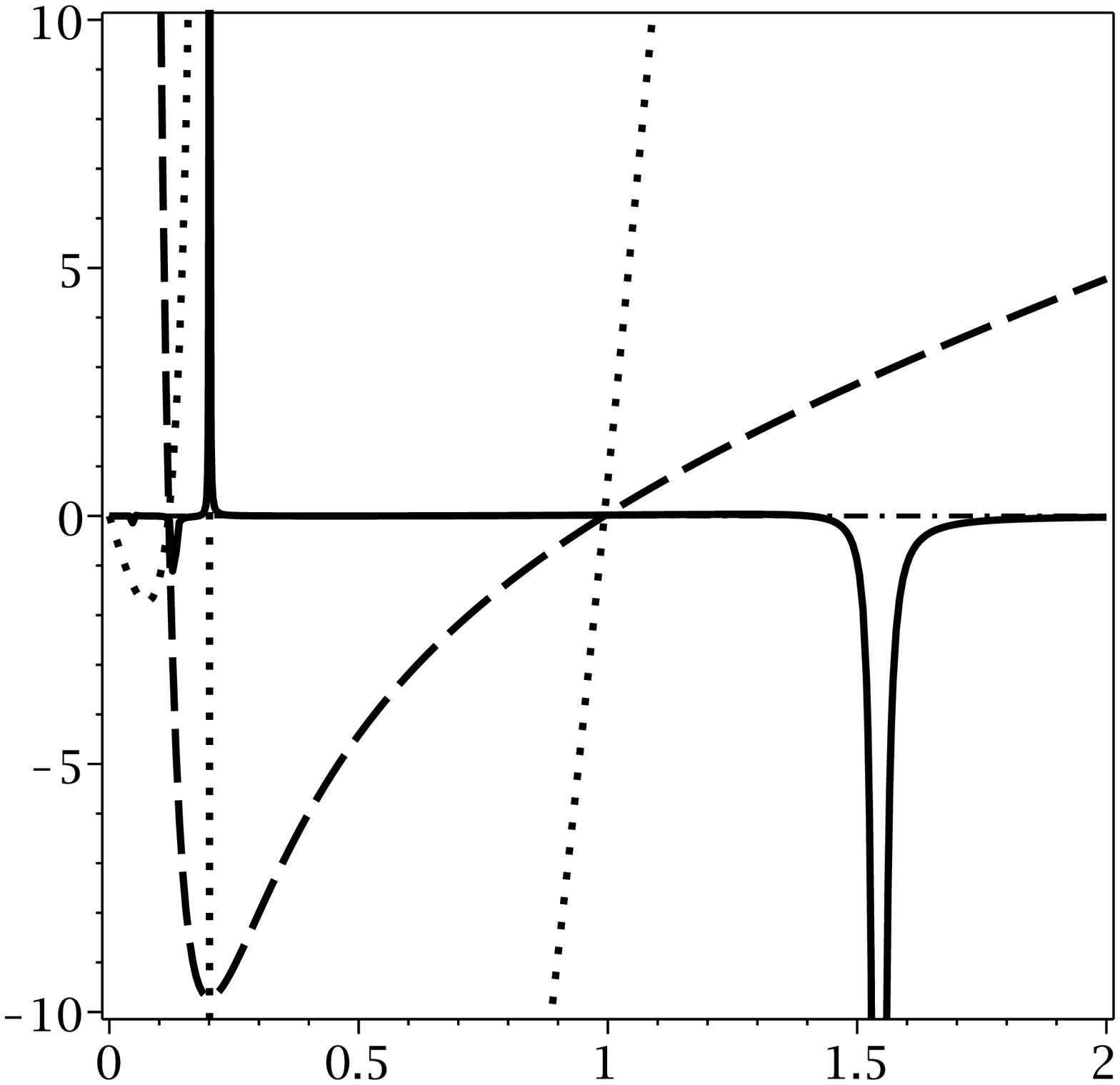} & \epsfxsize=7cm \epsffile{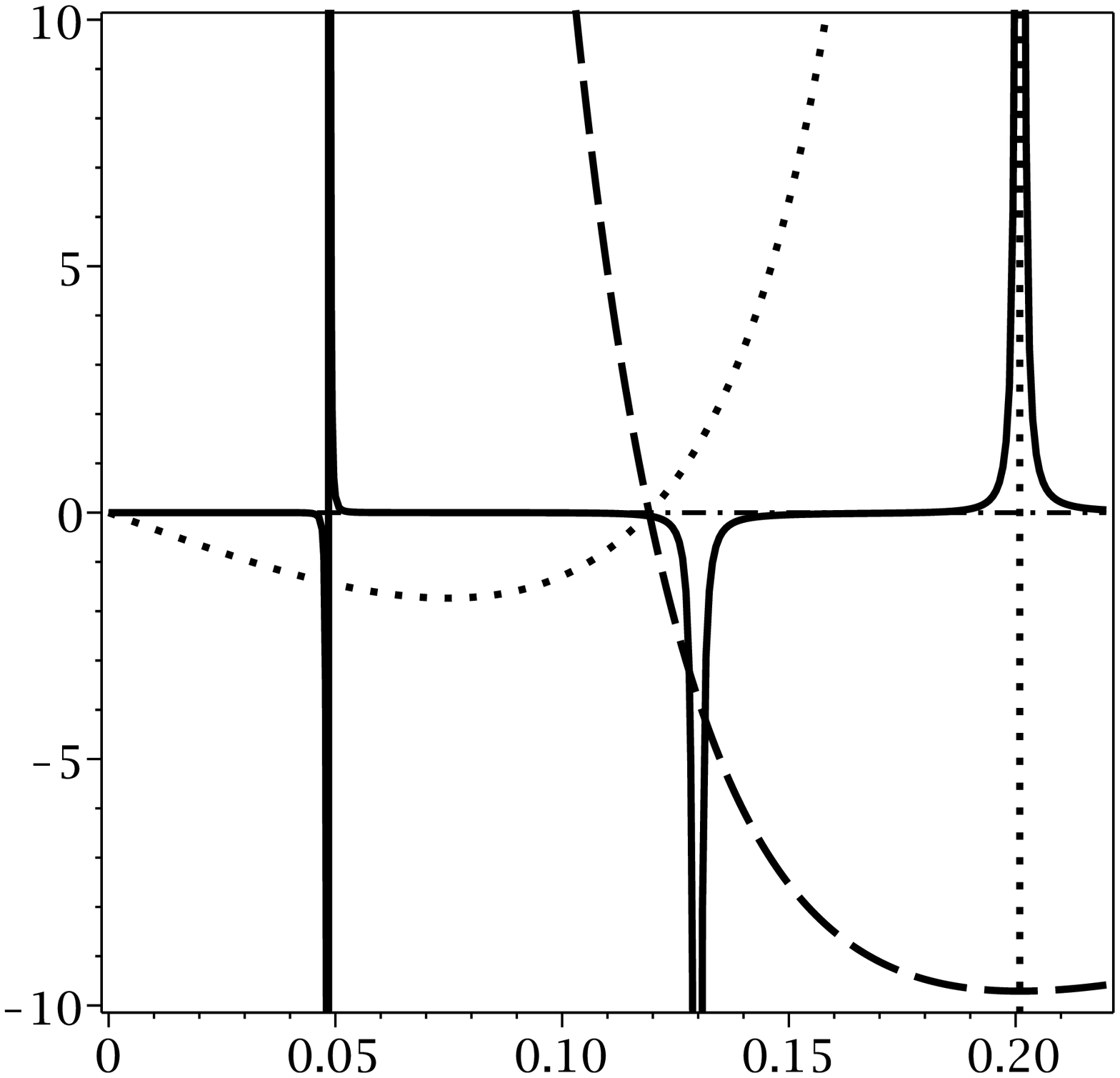}%
\end{array}
$%
\caption{Quevedo Ricci scalar case I (solid line), heat capacity (dotted
line) and temperature (dashed line) versus $S$ for $l=1$. \newline
\textbf{HNED model:} (up-left panel) and \textbf{SNED model:} (up-right
panel): $q=0.3$, $\protect\beta =1$. \newline
\textbf{CNED model:} (down-left panel) and \textbf{CNED model:} (down-right
panel): $q=1$, $\protect\alpha=0.007$ (different scales). }
\label{Fig3}
\end{figure}

\begin{figure}[tbp]
$%
\begin{array}{cc}
\epsfxsize=7cm \epsffile{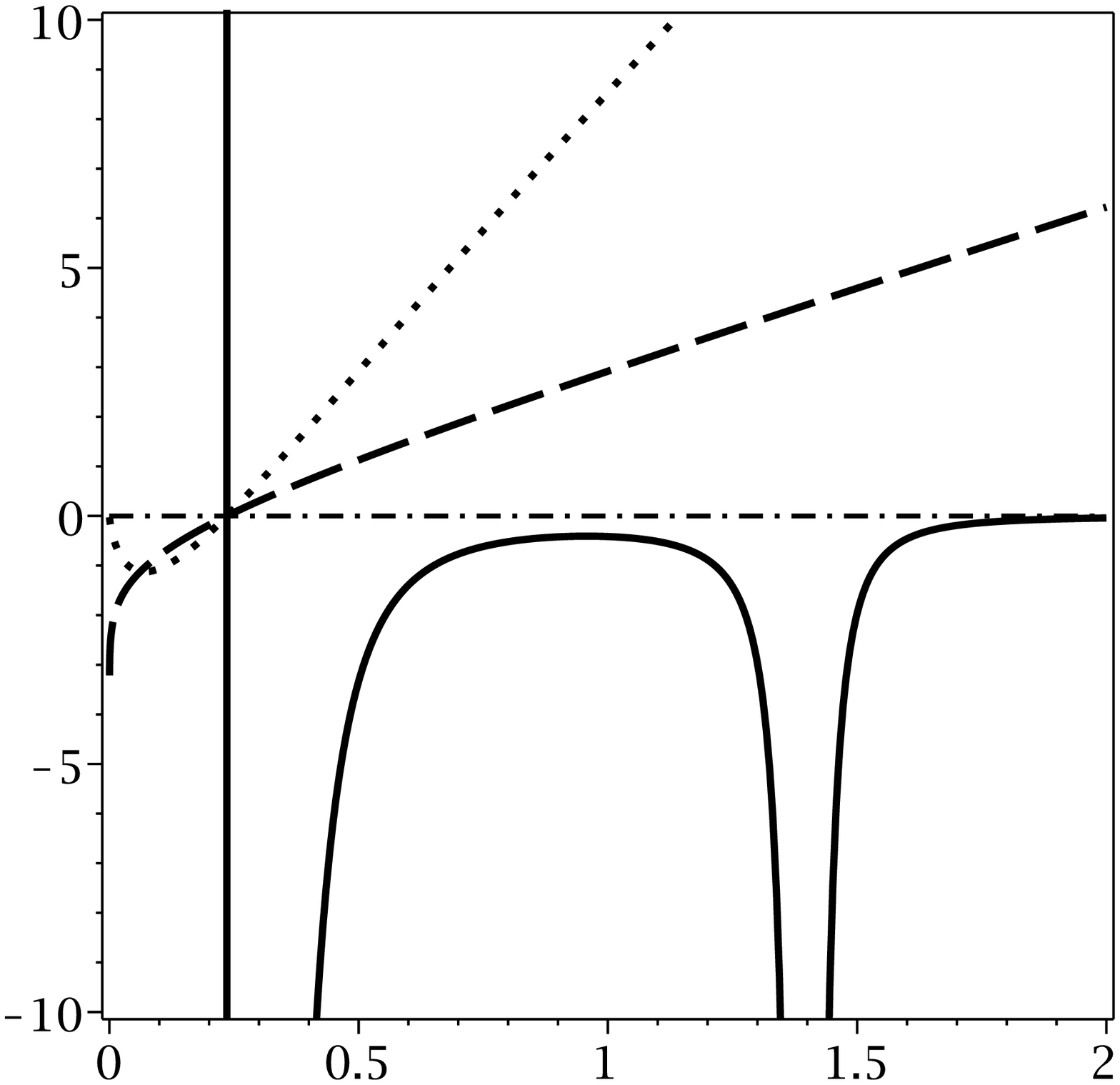} & \epsfxsize=7cm \epsffile{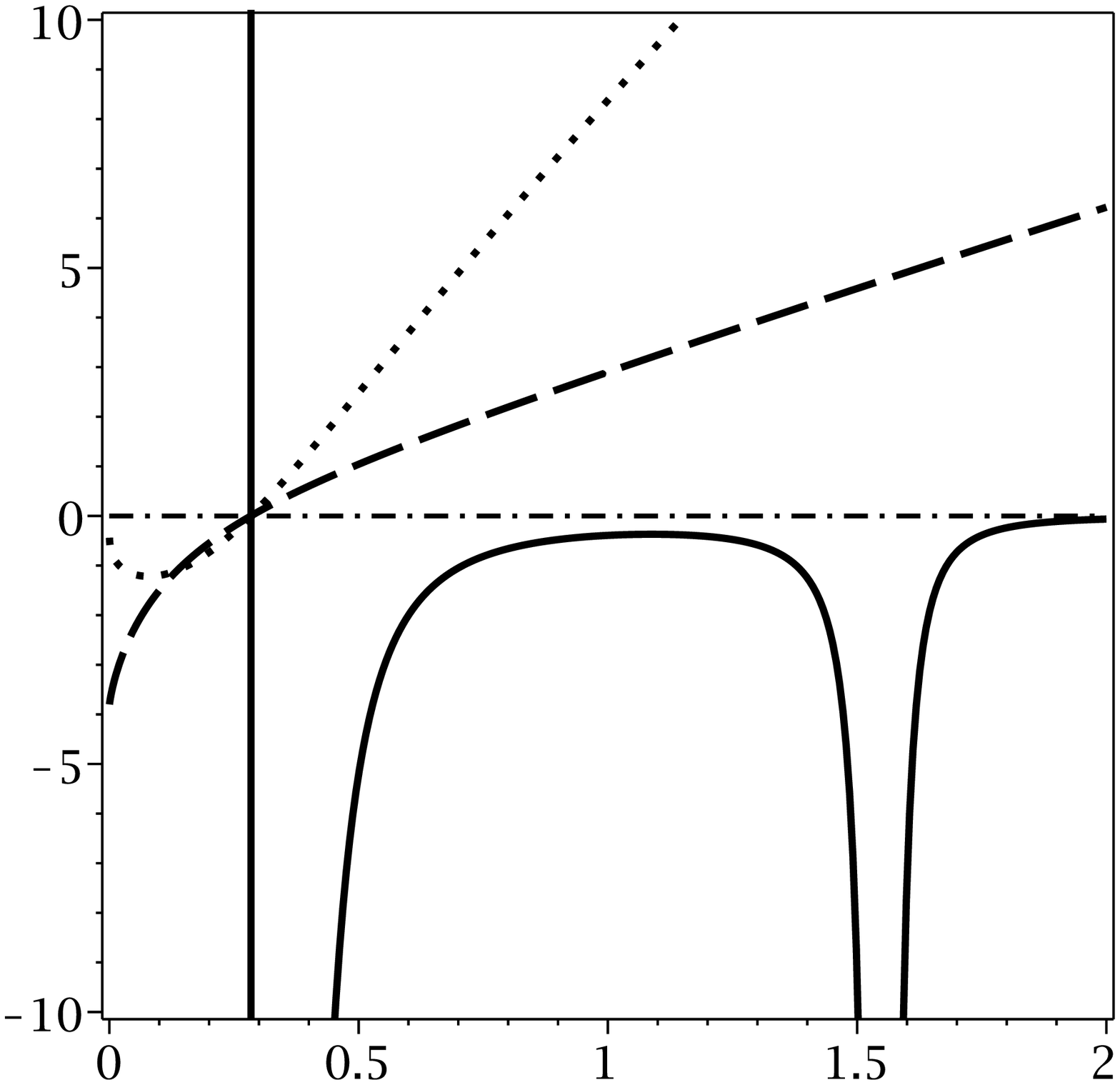} \\
\epsfxsize=7cm \epsffile{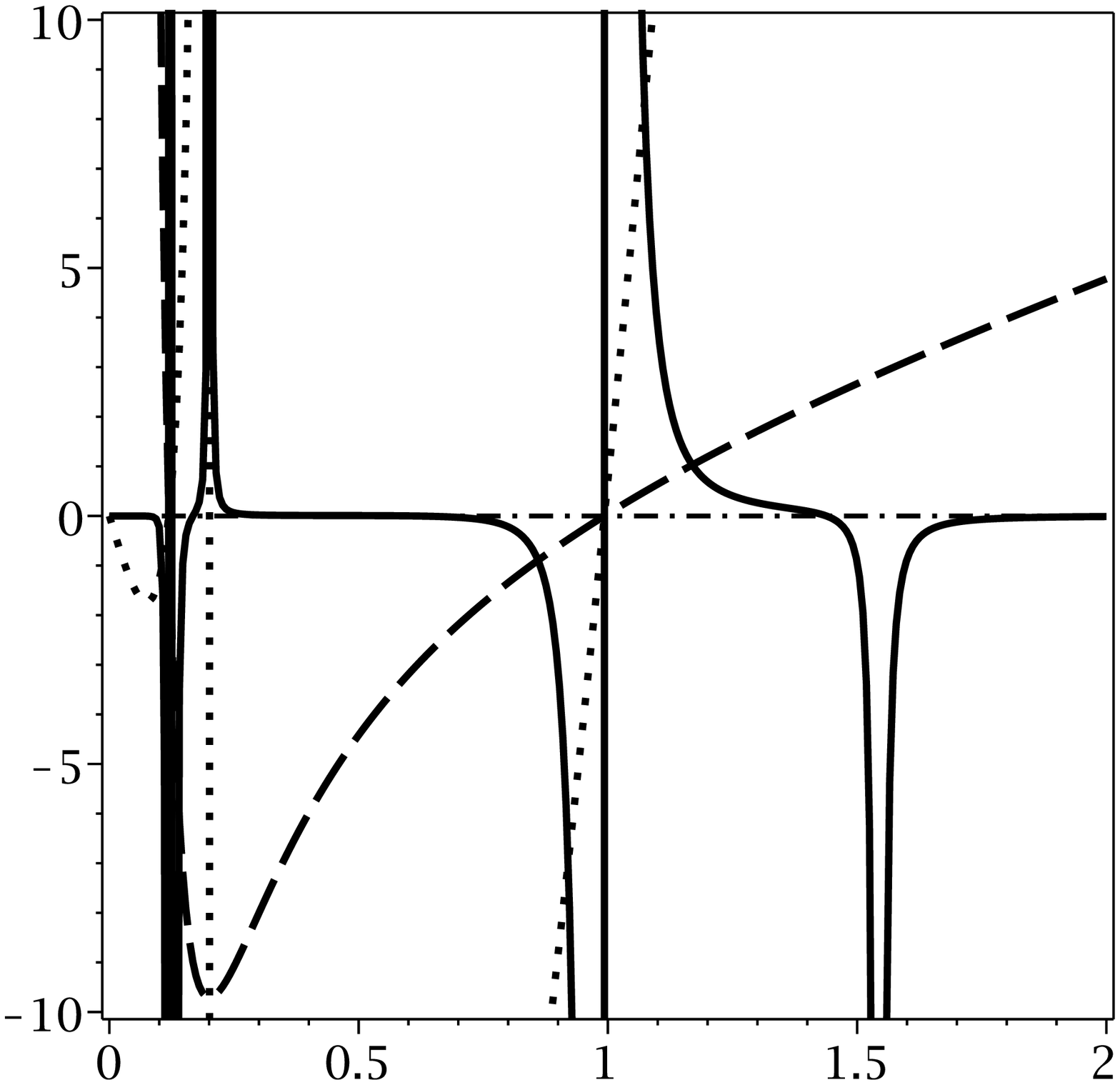} & \epsfxsize=7cm \epsffile{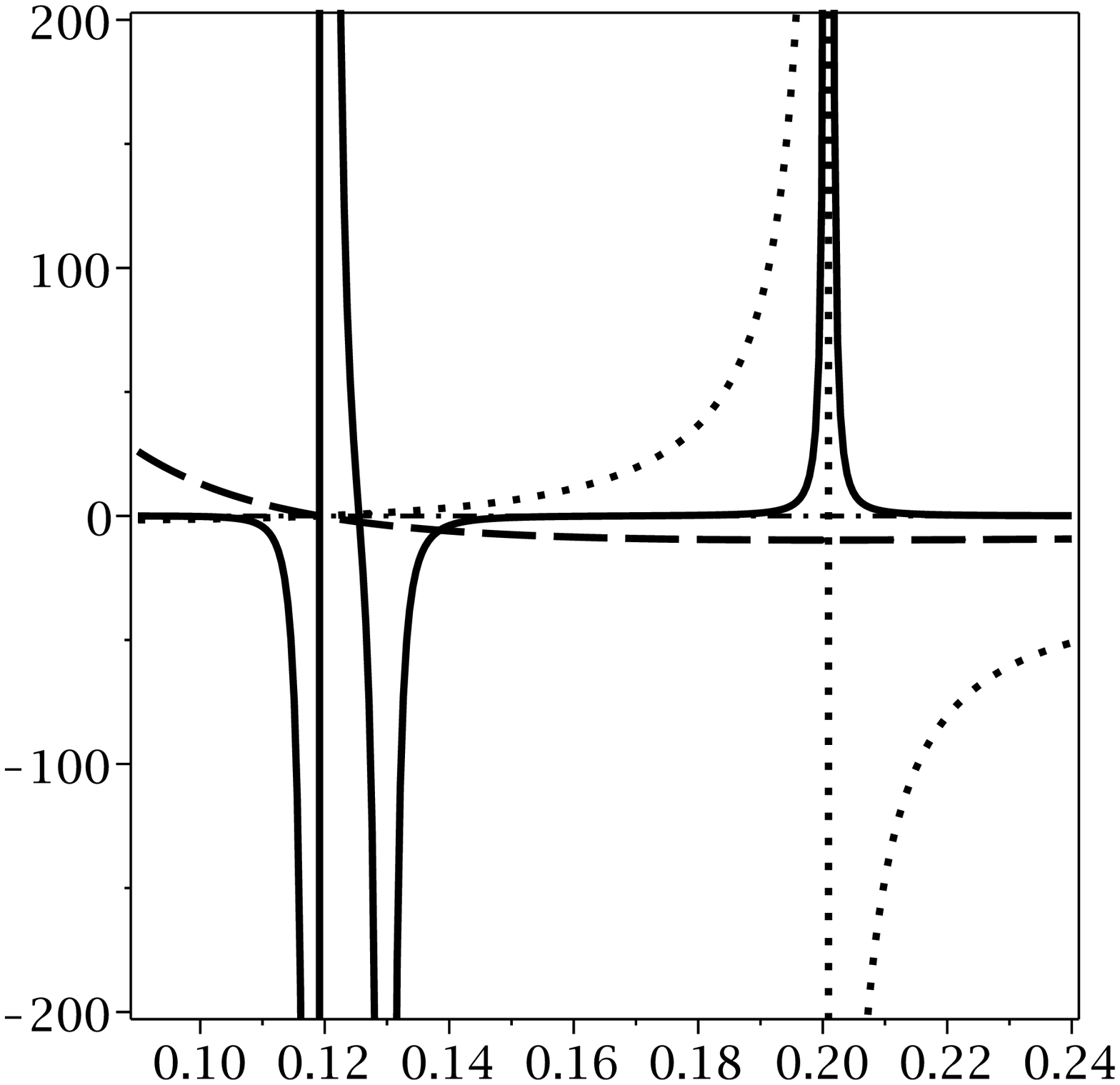}%
\end{array}
$%
\caption{Quevedo Ricci scalar case II (solid line), heat capacity (dotted
line) and temperature (dashed line) versus $S$ for $l=1$. \newline
\textbf{HNED model:} (up-left panel) and \textbf{SNED model:} (up-right
panel): $q=0.3$, $\protect\beta =1$. \newline
\textbf{CNED model:} (down-left panel) and \textbf{CNED model:} (down-right
panel): $q=1$, $\protect\alpha=0.007$ (different scales). }
\label{Fig4}
\end{figure}

\begin{figure}[tbp]
$%
\begin{array}{cc}
\epsfxsize=7cm \epsffile{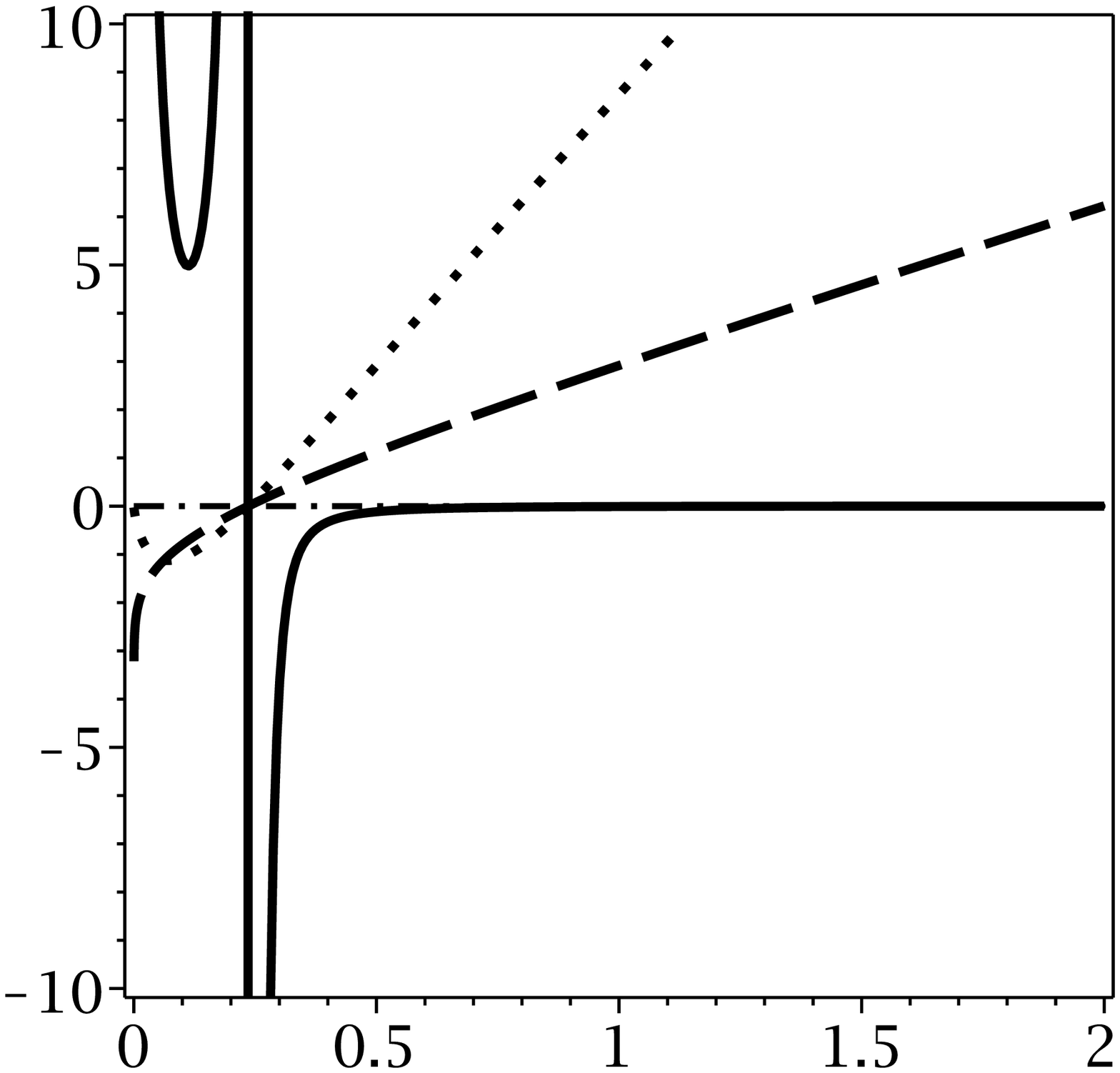} & \epsfxsize=7cm \epsffile{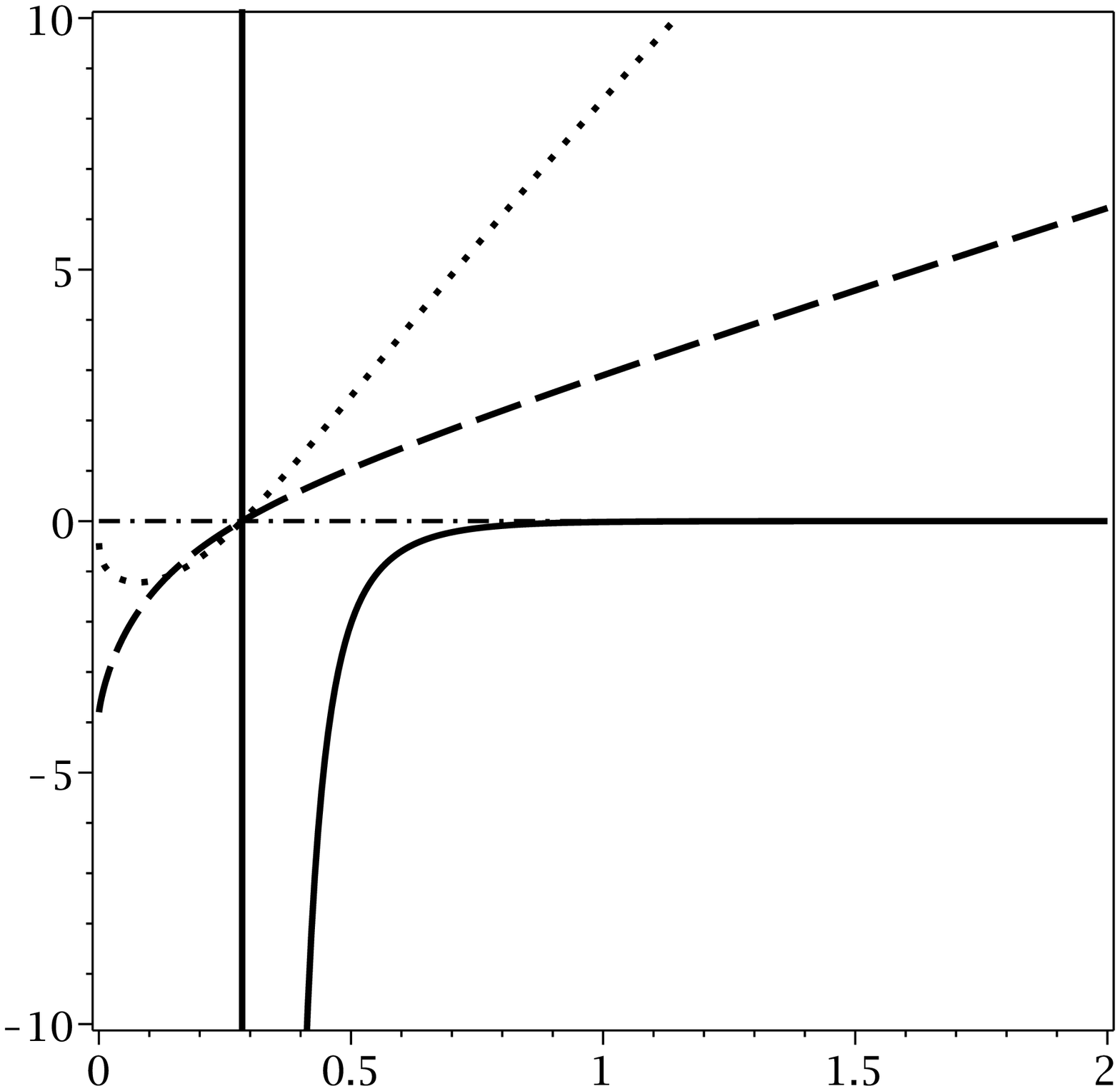}
\\
\epsfxsize=7cm \epsffile{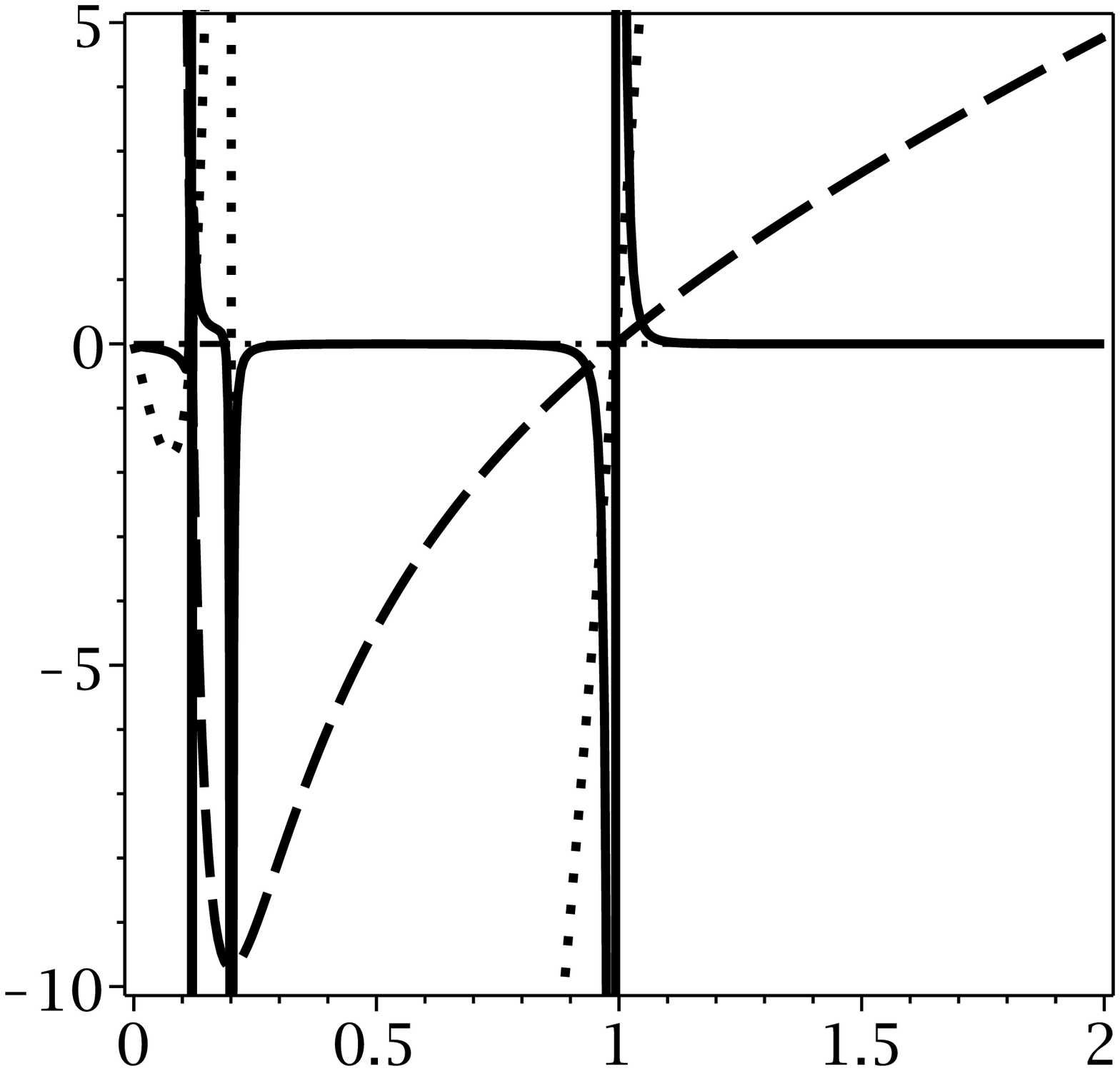} & \epsfxsize=7cm \epsffile{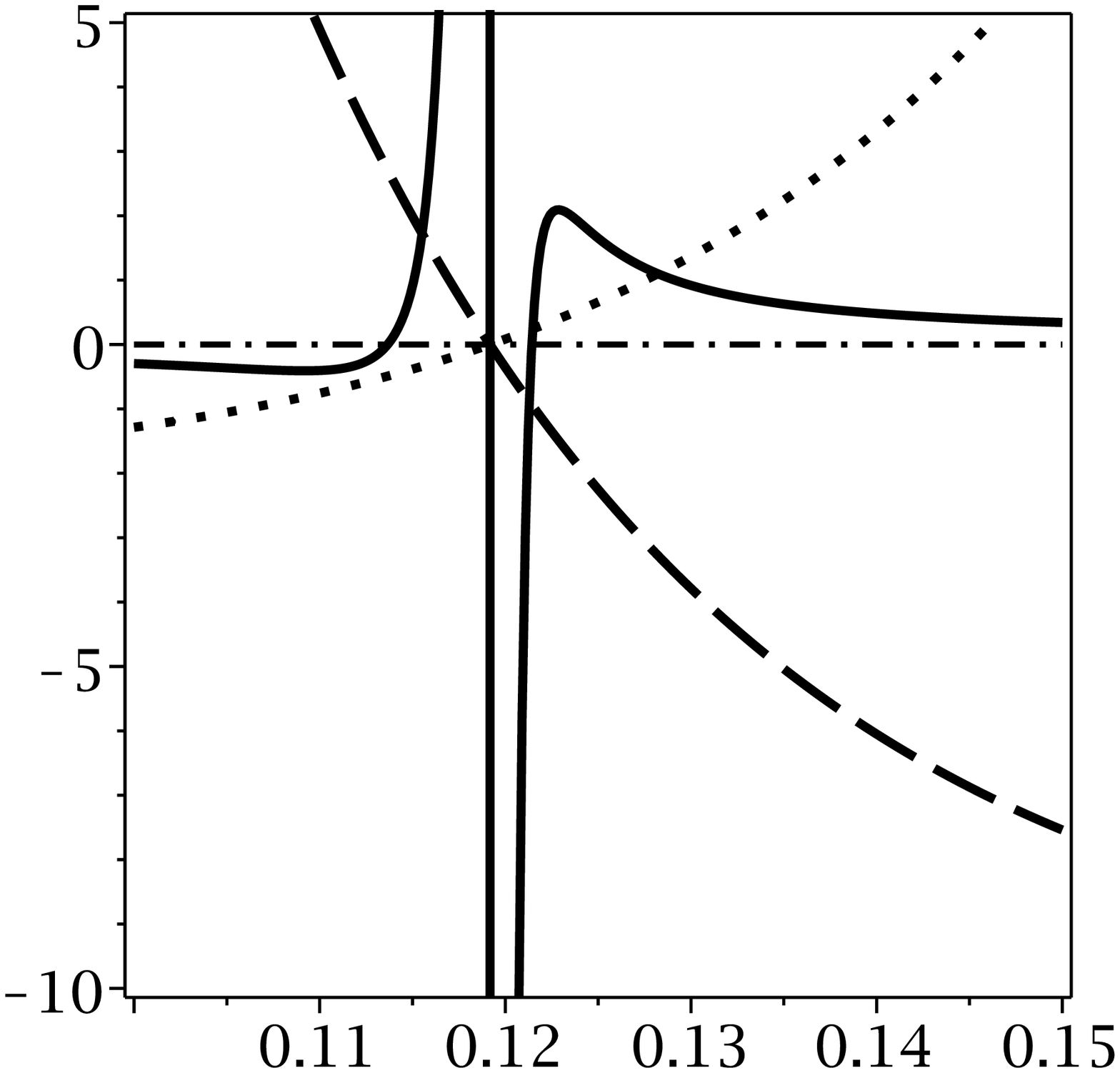}%
\end{array}
$%
\caption{HPEM Ricci scalar (solid line), heat capacity (dotted line) and
temperature (dashed line) versus $S$ for $l=1$. \newline
\textbf{HNED model:} (up-left panel) and \textbf{SNED model:} (up-right
panel): $q=0.3$, $\protect\beta =1$. \newline
\textbf{CNED model:} (down-left panel) and \textbf{CNED model:} (down-right
panel): $q=1$, $\protect\alpha=0.007$ (different scales). }
\label{Fig5}
\end{figure}

\subsubsection{Weinhold and Ruppeiner metrics}

The Weinhold metric was given in \cite{Weinhold1,Weinhold2}
\begin{equation}
dS_{W}^{2}=g_{ab}^{W}dX^{a}dX^{b},  \label{Wein}
\end{equation}
where $g_{ab}^{W}=\partial ^{2}M\left( X^{c}\right) /\partial X^{a}\partial
X^{b}$ and also $X^{a}\equiv X^{a}\left( S,N^{i}\right) $, where $N^{i}$
denotes other extensive variables of the system. In case of Weinhold
approach, one is considering the mass of the system as potential, other
parameters such as entropy and electric charge as extensive parameters and
related quantities such as temperature and electric potential as intensive
parameters.

The Ruppeiner metric was defined as \cite{Ruppeiner1,Ruppeiner2}
\begin{equation}
dS_{R}^{2}=g_{ab}^{R}dX^{a}dX^{b},  \label{Rupp}
\end{equation}
where $g_{ab}^{R}=-\partial ^{2}S\left( X^{c}\right) /\partial X^{a}\partial
X^{b}$ and $X^{a}\equiv X^{a}\left( M,N^{i}\right) $. In this case the
thermodynamical potential is entropy. It is worthwhile to mention that
according to the proposal of the Quevedo, these two approaches are related
to each other by a Legendre transformation \cite{WRconf}.

Taking into account thermodynamical metrics of Weinhold and
Ruppeiner, one can obtain their Ricci scalars. Since we would like
to investigate divergence points of TRS, $\mathcal{R}$, we focus
on its denominator ($D\mathcal{R}$). One finds
\begin{eqnarray}
D\mathcal{R}^{Win} &=&M^{2}\left( M_{SS}M_{QQ}-M_{SQ}^{2}\right) ^{2},
\label{R-Win} \\
D\mathcal{R}^{Rup} &=&TM^{2}\left( M_{SS}M_{QQ}-M_{SQ}^{2}\right) ^{2},
\label{R-Rup}
\end{eqnarray}%
where $M_{k}=\frac{\partial M}{\partial k}$ and $M_{kj}=\frac{\partial ^{2}M%
}{\partial k\partial j}$.

\subsubsection{The Quevedo metrics}

The Quevedo metrics have two kinds with the following forms \cite%
{Quevedo1,Quevedo2}
\begin{equation}
dS_{R}^{2}=g_{ab}^{Q}dX^{a}dX^{b},  \label{Quev}
\end{equation}%
where $g_{ab}^{Q}$ is%
\begin{equation}
g_{ab}^{Q}=\Upsilon \left(
\begin{array}{cc}
-M_{SS} & 0 \\
0 & M_{QQ}%
\end{array}%
\right) ,  \label{gQ}
\end{equation}%
with%
\[
\Upsilon =\left\{
\begin{array}{cc}
SM_{S}+QM_{Q}, & Case\text{ }I \\
SM_{S}, & Case\text{ }II%
\end{array}%
\right. .
\]%
Taking into account Quevedo metrics, one can find that their
related (denomerator of) Ricci scalars can be written as
\begin{eqnarray}
D\mathcal{R}^{Q-I} &=&M_{SS}^{2}M_{QQ}^{2}\left( SM_{S}+QM_{Q}\right) ^{3},
\label{R-QI} \\
D\mathcal{R}^{Q-II} &=&S^{3}M_{S}^{3}M_{SS}^{2}M_{QQ}^{2},  \label{R-QII}
\end{eqnarray}

\subsubsection{HPEM metric}

In order to avoid any extra divergencies in TRS which may not
coincide with phase transitions of the type one and two, and also
ensure that all the divergencies of the TRS coincide with phase
transition points of the both types, HPEM metric was introduced
\cite{HPEM}
\begin{equation}
g_{ab}=S\frac{M_{S}}{M_{QQ}^{3}}\left(
\begin{array}{cc}
-M_{SS} & 0 \\
0 & M_{QQ}%
\end{array}%
\right) .  \label{HPEM}
\end{equation}

In this case we have considered the total mass as thermodynamical
potential, entropy and electric charge as extensive parameters.
Calculations show that denominator of TRS leads to
\begin{equation}
D\mathcal{R}^{HPEM}=S^{3}M_{S}^{3}M_{SS}^{2}.  \label{R-HPEM}
\end{equation}

\section{The results of various approaches}

Here, we investigate phase transitions of black holes using
geometrothermodynamics. For this purpose, we used thermodynamical metrics
introduced in previous section for the black holes solutions obtained in the
section \ref{Sol}.

For Weinhold metric, none of divergencies of the Ricci scalar
coincide with roots of the heat capacity in every theories of NED
models that we have considered in this paper. On the other hand,
one of the divergencies of TRS and divergence point of the heat
capacity in CNED theory, coincide with each other. It is notable
that, in cases of the HNED and SNED theories, there is one
divergence point for TRS (up panels of Figs. \ref{Fig1}) whereas
for CNED, there are two divergencies (down panels of Figs.
\ref{Fig1}).

In case of Ruppeiner metric, for HNED and SNED models, there is a
root for heat capacity in which Ricci scalar of the Ruppeiner
metric has a divergency. But there is also another divergence
point for Ricci scalar which does not coincide with any phase
transition point (up panels of Figs. \ref{Fig2}). Therefore, there
is an extra divergence point. In case of the CNED model, two roots
and one divergence point are observed for heat capacity in which
Ricci scalar has related divergencies. In addition to these
divergence points, one extra divergence point is also observed
which is not related to any phase transition point (down panels of
Figs. \ref{Fig2}).

As for Quevedo metrics, for case I, similar behavior as Weinhold
is observed for all three theories of NED (Fig. \ref{Fig3}). On
the other hand, in case of the other metric of Quevedo, two
divergence points for Ricci scalar were observed for HNED and SNED
theories. One of these divergence points coincides with root of
the heat capacity for these two nonlinear theories whereas the
other one does not (up panels of Figs. \ref{Fig4}). In case of
CNED theory, all the divergence points of TRS coincide with phase
transition points except one (down panels of Figs. \ref{Fig4}). In
other words, Quevedo's metric predicts an extra divergence point,
corresponding to the equation $M_{QQ} = 0$, which is not predicted
in classical black hole thermodynamics.

It is evident that in case of HPEM, all types of phase transition
points of heat capacity coincide with divergencies of TRS of HPEM
method (Figs. \ref{Fig5}). In other words, independent of the
nonlinear theory under consideration, the HPEM method provide a
machinery in which no extra divergency for TRS is observed and
divergence points of TRS and phase transition points coincide.
Another interesting and important property of the HPEM method is
the behavior of TRS near divergence point for different types of
phase transition. As one can see, the signature and behavior of
TRS near divergence point for phase transition type one and two
are different. Therefore, independent of studying the heat
capacity, one can distinguish these types of phase transition from
one another only by studying the behavior of TRS.


\section{Closing Remarks}

In this paper, we have considered BTZ black holes, in presence of three
models of NED. We studied stability and phase transitions related to the
heat capacity of the mentioned black holes. Next, we employed geometrical
approach to study the thermodynamical behavior of the system. In other
words, we have studied phase transitions of the system through Weinhold,
Ruppeiner and Quevedo methods. Also, we used the recently proposed approach
to study geometrical thermodynamics.

We found that the Weinhold and Ruppeiner metrics for studying
these BTZ solutions fail to provide a suitable result. In
addition, the divergence points of the Quevedo TRS were not
completely matched with the phase transition points of the heat
capacity results. In other words, in these approaches, the
existence of extra divergencies were observed which were not
related to any phase transition point in the classical
thermodynamics. In some of these approaches no divergency of TRS
coincided with phase transition points. In order to obtain a
consistent results with the classical thermodynamic consequences
(the heat capacity), we employed a new thermodynamical metric. In
this approach, all the divergencies of TRS coincided with phase
transition points. In other words, roots and divergence points of
the heat capacity of the BTZ black holes in presence of each
nonlinear models matched with divergencies of TRS of this metric.

Also, we found that in case of HNED and SNED theories, there is no
divergency for heat capacity. It means that, like Maxwell theory, these two
theories have no second type phase transition. These two nonlinear theories
of electrodynamics, preserved the characteristic behavior of the Maxwell
theory in case of heat capacity. On other hand, for CNED model, two roots
and one divergence point was found for the heat capacity. In other words,
due to contribution of the nonlinear electromagnetic field, heat capacity
enjoys the existence of one more phase transition point of type one and a
phase transition of the type two. In essence this theory is a generalization
of the Maxwell theory. But this generalization added another property to
heat capacity that was not observed for the Maxwell theory.

Finally it is worthwhile to mention a comment related to Legendre
invariancy. It was shown that \cite{LI} the Legendre invariance
alone is not sufficient to guarantee a unique description of
thermodynamical metrics in terms of their curvatures. In addition
to Legendre invariancy, one needs to demand curvature invariancy
under a change of representation. Therefore, it will be worthwhile
to investigate both Legendre and curvature invariancies. In
addition, it will be interesting to think about the fundamental
relation between the following two issues: (I) Agreement of
thermodynamical curvature results with usual thermodynamical
approaches (such as the heat capacity); (II) Curvature invariancy
in addition to the Legendre invariancy. It is also worthwhile to
probe the fundamentality of cases (I) and (II) to find that
considering which one may leads to satisfy another one. Although
first issue has been investigated for special cases \cite{LI}, the
second one has been remained unanalyzed yet. We may address them
in an independent work in the future.

\begin{center}
\textbf{Conflict of Interests }
\end{center}

The authors declare that there is no conflict of interests regarding the
publication of this paper.

\begin{acknowledgements}
We would like to thank the anonymous referee for valuable
suggestions. We also thank the Shiraz University Research Council.
This work has been supported financially by the Research Institute
for Astronomy and Astrophysics of Maragha, Iran.
\end{acknowledgements}

\end{document}